%% file: main.tex
\documentclass[10pt, a4paper,twocolumn]{article}

\usepackage[utf8]{inputenc}
\usepackage[T1]{fontenc}
\usepackage{fontawesome5}

\usepackage[hidelinks, colorlinks=false]{hyperref}
\usepackage{bookmark}
\usepackage{hyperxmp}
\usepackage{romannum}

\usepackage{amsmath}
\usepackage{amssymb}
\usepackage{amsthm}
\usepackage{mathtools}

\usepackage[english]{cleveref}

\usepackage[top=2cm, bottom=2cm, left=1.5cm, right=1.5cm, columnsep=0.6cm]{geometry}
\usepackage{graphicx}

\usepackage{tikz}
\usetikzlibrary{calc}
\usepackage{subcaption}
\usepackage{enumitem}
\usepackage{listings}
\usepackage{titlesec}
\usepackage{parskip}
\usepackage{booktabs}
\usepackage{tabularx}
\usepackage{braket}
\usepackage{newunicodechar}
\usepackage{setspace}
\usepackage[dvipsnames]{xcolor}
\usepackage[many]{tcolorbox}
\tcbuselibrary{theorems, breakable}
\usepackage[titles]{tocloft}

\usepackage{academicons}
\definecolor{orcidlogocol}{HTML}{A6CE39}

\usepackage[
  backend=biber,
  style=numeric-comp,
  natbib=true,
  sorting=nyt,
  maxbibnames=99,
  maxcitenames=2,
  giveninits=true,
  uniquename=init,
  doi=true,
  url=false,
  isbn=false,
  eprint=false
]{biblatex}
\input{helper_user_definitions}

\title{From Classical to Quantum-Mechanical Data Assimilation: \\
       \vspace{-1em} A Comparison between DATO and QMDA \vspace{0.5em}}

\author{%
    \centering{
        \href{https://orcid.org/0009-0006-2126-4816}{\textcolor{orcidlogocol}{\aiOrcid}} Emanuele Donno\affmark{a,b} \;
        \href{https://orcid.org/0000-0001-5083-2671}{\textcolor{orcidlogocol}{\aiOrcid}} Giovanni Conti\affmark{c} \;
        \href{https://orcid.org/0000-0003-4451-3898}{\textcolor{orcidlogocol}{\aiOrcid}} Paolo Oddo\affmark{d} \;
        \href{https://orcid.org/0000-0001-7777-8935}{\textcolor{orcidlogocol}{\aiOrcid}} Silvio Gualdi\affmark{c} \;
        \href{https://orcid.org/0000-0001-9387-9277}{\textcolor{orcidlogocol}{\aiOrcid}} Luca Mainetti\affmark{a} \;
        \href{https://orcid.org/0000-0001-5902-6983}{\textcolor{orcidlogocol}{\aiOrcid}} Giovanni Aloisio\affmark{a,b}
        \\[0.6em]
        \normalsize\small
        {\raggedright
            \orcid{a}Department of Innovation Engineering - University of Salento, Lecce, Italy\\
            \orcid{b}CMCC Foundation - Euro-Mediterranean Center on Climate Change, Lecce, Italy\\
            \orcid{c}CMCC Foundation - Euro-Mediterranean Center on Climate Change, Bologna, Italy\\
            \orcid{d}Department of Physics and Astronomy - Alma Mater Studiorum University of Bologna, Bologna, Italy\\
        }
    }
}
\date{}

\begin{document}

\pagenumbering{arabic}

\twocolumn[{%
    \maketitle
    \begin{abstract}
    \input{sec_0_abstract.tex}
    \end{abstract}
    \vspace{1.5em}
}]

\input{sec_1_intro.tex}
\input{sec_2_notations_and_conventions.tex}

\input{sec_3_framework_DATO.tex}
\input{sec_4_framework_QMDA.tex}
\input{sec_5_comparison.tex}
\input{sec_6_lorenz_63.tex}
\input{sec_7_conclusions.tex}
\input{sec_8_CRediT.tex}
\newpage

\printbibliography
\end{document}

%% file: helper_user_definitions.tex
\newunicodechar{—}{---}

\newcommand{\affmark}[1]{\textsuperscript{{\textit{#1}}}}
\newcommand{\orcid}[1]{\textsuperscript{\,\textit{#1}}}

\titleformat{\section}{\color{black}\Large\bfseries\scshape}{\thesection.}{0.5em}{}
\titlespacing*{\section}{0pt}{1.4ex plus 0.5ex}{0.8ex plus 0.5ex}
\titleformat{\subsection}{\color{black}\large\bfseries}{\thesubsection.}{0.5em}{}
\titlespacing*{\subsection}{0pt}{1.0ex plus 0.5ex}{0.5ex}
\titleformat{\subsubsection}{\color{black}\bfseries}{\thesubsubsection.}{0.5em}{}
\titlespacing*{\subsubsection}{0pt}{1.0ex plus 0.5ex}{0.5ex}

\makeatletter
\renewcommand{\maketitle}{%
  \begin{center}
	\rule{\linewidth}{2pt}
	\vspace{2em}
    {\LARGE\bfseries\color{black} \@title \par}
    \vspace{0.6em}
    {\small \@author \par}
    \vspace{0.3em}
    {\small \@date \par}
	\vspace{1em}
	\rule{\linewidth}{2pt}
	\vspace{1em}
  \end{center}
}
\makeatother

%% file: sec_0_abstract.tex
Data assimilation provides a systematic framework for combining dynamical models with partial and noisy observations to infer the evolving state of a system. In this work, we undertake a comparative study of \emph{Data Assimilation with Transfer Operators} (DATO) and \emph{Quantum Mechanical Data Assimilation} (QMDA), focusing on their mathematical formulation, algorithmic structure, and empirical performance. Both methods are first cast within a common operator-theoretic framework, which makes it possible to compare, on a unified basis, their representations of uncertainty, forecast propagation, and assimilation updates. We then analyse their principal similarities and differences with respect to state-space structure, update mechanisms, structural preservation properties, and computational cost. To complement the theoretical analysis, we assess both approaches on benchmark dynamical systems across a range of observational settings, including noisy, sparse, and partially observed regimes. Our results show that, despite their shared operator-theoretic motivation, DATO and QMDA embody substantially different assimilation paradigms, leading to distinct advantages and limitations in terms of interpretability, robustness, and scalability. The present study helps delineate the regimes in which each framework is most effective and offers broader insight into the design of operator-based methodologies for data assimilation.

%% file: sec_1_intro.tex
\section{Introduction}
\label{sec:introduzione}

Numerical models of geophysical, climatic and engineering systems are necessarily imperfect. Discretisation errors, unresolved sub-grid processes and parametric uncertainty cause simulated trajectories to depart from the underlying physical state, while chaotic sensitivity to initial conditions can rapidly amplify this mismatch. Observations mitigate this deficiency, but they are sparse, irregular in space and time, and affected by instrumental noise. Data assimilation (DA) provides the probabilistic framework for combining numerical forecasts with observations in order to reconstruct the state of a dynamical system and quantify the associated uncertainty \citep{asch2016da,evensen2022fundamentals}. The same formalism also extends naturally to parameter estimation and to the identification of uncertain forcings or control inputs.

In its standard Bayesian formulation, DA is described by a stochastic model equation and an observation equation,
\[
 x_{k+1}=M[x_k]+\eta_k, \qquad
 y_k=H[x_k]+\varepsilon_k,
\]
where $x_k\in\mathbb{R}^n$ is the model state at time $t_k$, $y_k\in\mathbb{R}^p$ is the observation vector, $M$ is the model operator, $H$ maps the state space into the observation space, and $\eta_k$, $\varepsilon_k$ denote model and observation errors.

By combining the stochastic model equation, which determines the prior $\rho(x_{k} \mid y_{1:k-1})$ through marginalisation over the previous posterior, with the observation equation, which determines the likelihood $\rho(y_{k} \mid x_{k})$, Bayes' theorem yields the posterior $\rho(x_{k} \mid y_{1:k}) \propto \rho(y_{k} \mid x_{k}) \, \rho(x_k \mid y_{1:k-1})$.

At each assimilation cycle the target is the posterior density $\rho(x_k\mid y_{1:k})$, obtained by alternating a \emph{forecast} step, which advances the prior through the model dynamics, and an \emph{analysis} step, which updates the prior with the new observations through Bayes' theorem. This predictor--corrector structure is common to sequential DA schemes.

Operational DA is dominated by two broad algorithmic families. \textit{Variational methods}, such as 3D-Var and 4D-Var, formulate the analysis as the minimisation of a cost functional over an assimilation window, balancing distance from a background state against distance from observations \citep{lorenc1986,sasaki1970,courtier1994fourdvar}. They can be highly effective, but generally require the construction and maintenance of an adjoint model. \textit{Sequential statistical methods} originate with the Kalman Filter \citep{kalman1960,jazwinski1970stochastic}. The Ensemble Kalman Filter replaces explicit covariance propagation with an ensemble of model integrations \citep{evensen1994enkf,evensen2003enkf,anderson2001eakf}, while Particle Filters approximate the full posterior by weighted samples \citep{vanleeuwen2009pf,vanleeuwen2019pf}.

These methods face three structural limitations. First, every assimilation cycle requires explicit model integration, and often many such integrations; in high-dimensional geophysical systems, where $n$ may reach $10^7$--$10^9$, this dominates the computational budget \citep{law2015da}. Second, Kalman-type and variational schemes are optimal only under assumptions close to linearity and Gaussianity, so they can become biased in strongly nonlinear or multimodal regimes. Third, Particle Filters avoid Gaussian assumptions but suffer from weight degeneracy: the required ensemble size may grow exponentially with dimension unless localisation, tempering or hybrid corrections are introduced \citep{snyder2008obstacles,vanleeuwen2019pf}. Moreover, all these approaches presuppose access to an explicit or repeatedly executable model $\mathcal{M}$, which is problematic when the dynamics are only partially known or available only through expensive black-box solvers.

A complementary route comes from operator-theoretic descriptions of dynamics. Given a discrete map $x_{k+1}=M[x_k]$ on a state space $A\subset\mathbb{R}^n$ with invariant measure $\mu$, the Koopman operator \citep{koopman1931,koopmanvonneumann1932} acts linearly on observables $g\in L^2(A,\mu)$ by
\begin{equation}
 (\mathcal{K}_{\Delta t}g)(x)=g(M[x]),
 \label{eq:koopman}
\end{equation}
and its adjoint, the Perron--Frobenius operator, evolves probability densities according to
\begin{equation}
 \rho_{k+1}(x)=(\mathcal{P}_{\Delta t}\rho_k)(x).
 \label{eq:pf}
\end{equation}
Although the underlying map $M$ may be nonlinear, these operators are linear on infinite-dimensional function spaces. Their spectra encode coherent structures and intrinsic time scales of the dynamics. Finite-dimensional, data-driven approximations such as Dynamic Mode Decomposition, EDMD and kernel EDMD make this perspective computationally usable \citep{schmid2010dmd,williams2015edmd,klus2018ddmr,klus2019eigen}: once a surrogate operator is learned, forecasting can be reduced from numerical integration of an ODE or PDE to algebraic operations on spectral coefficients. Exploiting the eigenfunction $\phi_{j}^{P}$ and $\phi_{j}^{K}$ of the two operators above, togheter with their eigenvalues $\lambda$, it is possible to provide the forecast of the system and its correction.

This work focuses on two recent operator-theoretic DA frameworks. \textbf{Data Assimilation with Transfer Operators} (DATO) \citep{conti2025dato} estimates Koopman and Perron--Frobenius operators through kernel EDMD with a Gaussian RBF kernel and Tikhonov regularisation. The state is represented by coefficients $\xi\in\mathbb{R}^S$ in a truncated Perron--Frobenius eigenfunction basis, and forecast propagation is diagonal:
\begin{equation}
 \xi_j^{(k+1)}=\lambda_j^q\xi_j^{(k)},\qquad j=1,\ldots,S,
\end{equation}
where $q$ is the number of model time steps between observations. The analysis applies a pointwise Gaussian likelihood and re-projects the posterior onto the spectral basis, producing a point estimate and posterior density values on the training set. The operator structure also yields closed-form expressions for diagnostics such as Observation Influence and Forecast Sensitivity to Observation Impact \citep{cardinali2004influence,cardinali2009monitoring,cardinali2014oi,cardinali2014impact}.

\textbf{Quantum Mechanical Data Assimilation} (QMDA) \citep{giannakis2019qmda} instead reformulates sequential DA using the language of Dirac--von Neumann quantum mechanics \citep{takhtajan2008qmm}. The state is a trace-class density operator $\hat{\rho}\in\mathbb{R}^{L\times L}$ on a truncated spectral subspace; observables are self-adjoint multiplication operators; forecast evolution is induced by the Koopman group; and the Bayesian update is realised through the projective dynamics of the von Neumann measurement postulate. The basis is built from a variable-bandwidth kernel with bistochastic normalisation, optionally combined with delay-coordinate maps when only partial observations are available \citep{berry2016vbkernel,coifman2013bistochastic,takens1981delay,sauer1991embedology}. Its output is a discrete probability distribution over bins of the observed quantity.

DATO and QMDA therefore share an operator-theoretic foundation but differ substantially in representation, propagation and output. DATO returns the posterior density evaluated at the training points, from which a point estimate $x^a$ is computed as the posterior mean; QMDA returns a discrete probability distribution over a partition of the observable range, and a state-space point estimate is not directly produced.

Both have been tested on the Lorenz--63 system \citep{lorenz1963},
\begin{equation}
\begin{cases*}
\dot{x}=\gamma(y-x),\\
\dot{y}=x(\omega-z)-y,\\
\dot{z}=xy-\beta z,
\end{cases*}
\qquad (\gamma,\omega,\beta)=(10,28,8/3),
\label{eq:l63}
\end{equation}
a low-dimensional but chaotic benchmark whose non-Gaussian invariant measure makes it a standard nonlinear DA test case. In the reference configurations, DATO uses $m=2{,}800$ training snapshots \citep{conti2025dato}, while QMDA uses $N=64{,}000$ training samples \citep{giannakis2019qmda}; despite the identical physical dimension $n=3$, these choices already lead to markedly different online costs.

To the best of the authors' knowledge, no systematic comparison of DATO and QMDA has yet been carried out, especially from the viewpoint of computational complexity. This work fills that gap by providing: (i) a unified asymptotic analysis of the offline and online phases of both frameworks; (ii) a direct comparison at equal size, separating structural differences from the dimensional choices of the reference experiments; (iii) the derivation of the break-even threshold $n^{*}=L^3/m$ as a quantitative selection criterion; and (iv) operational estimates on Lorenz--63, used to check the consistency of the asymptotic profiles with the reported experimental scales.

The remainder of the work is organised as follows. \Cref{sec:notazione} introduces notation and offline/online conventions. \Cref{sec:framework_dato} and \cref{sec:framework_qmda} present DATO and QMDA together with their cost analyses. \Cref{sec:confronto} compares the two frameworks and derives the break-even criterion. \Cref{sec:l63} applies the analysis to Lorenz--63, while \cref{sec:conclusioni} and \cref{sec:future_works} summarise the findings and outline future developments.

%% file: sec_2_notations_and_conventions.tex
\section{Notation and Conventions}
\label{sec:notazione}

All symbols used in the analysis and in the comparison are collected in the following table:

\begin{table}[htp!]
	\centering
	\renewcommand{\arraystretch}{1.2} \small
	\begin{tabularx}{\columnwidth}{@{}lX@{}}
		\toprule
		\textbf{Sym.} & \textbf{Scope \& Meaning} \\
		\midrule
		$m$   & \small{\textbf{DATO}}: Number of training snapshots; \\
		$S$   & \small{\textbf{DATO}}: Number of retained PF/Koopman eigenpairs; \\
		$\sigma$ & \small{\textbf{DATO}}: Gaussian-RBF kernel bandwidth; \small{\textbf{QMDA}}: variable bandwidth $\sigma_N$ from the kernel of \citet{berry2016vbkernel} \\
		$\epsilon$ & \small{\textbf{DATO}}: Tikhonov regulariser; \small{\textbf{QMDA}}: bandwidth scale of the variable-bandwidth kernel; \\
		\midrule
		$N$   & \small{\textbf{QMDA}}: Number of training samples; \\
		$S$   & \small{\textbf{QMDA}}: Cardinality of partition $\Xi$ of the observable range \\
		$L$   & \small{\textbf{QMDA}}: Spectral resolution (kernel basis);\\
		$d$   & \small{\textbf{QMDA}}: Data-space dimension; $d=n$ for full-state observations, $d=Q$ for delay-coordinate maps with $Q$ delays \\
		$r$   & \small{\textbf{QMDA}}: Number of neighbours retained in the sparse kernel approximation $\hat{\mathbf{G}}$ ($r\ll N$) \\
		\midrule
		$n$   & \small{\textbf{Both}}: State-space dimension, $\mathbf{x}\in \mathbb{R}^n$ \\
		$p$   & \small{\textbf{Both}}: Observation-space dimension, $\mathbf{y}\in\mathbb{R}^p$ \\
		$K$   & \small{\textbf{Both}}: Number of assimilation cycles \\
		$q$   & \small{\textbf{Both}}: Model steps between consecutive observations (physical parameter; e.g. $q=100$ in L63 setup of QMDA) \\
		\bottomrule
	\end{tabularx}
	\caption{Symbols used in the analysis. Scope ``Both'' marks shared symbols; otherwise framework-specific.}
	\label{tab:notazione}
\end{table}

\paragraph{Note on symbol overlap}
Three symbol clashes arise between the two reference papers and are resolved here as follows.

\begin{enumerate}
  \item In \citet{giannakis2019qmda} the letter $m$ denotes the data-space dimension. To avoid collision with the snapshot count $m$ of \citet{conti2025dato}, the QMDA data-space dimension is renamed $d$ throughout this work.
  
  \item The symbol $S$ is reused, with two distinct meanings: in DATO, $S$ is the number of retained Perron--Frobenius / Koopman eigenpairs (with $S \le m$); in QMDA, $S$ is the cardinality of the partition $\Xi=\{\Xi_0,\dots,\Xi_{S-1}\}$ of the range of the observable. The two values are unrelated and need not coincide; context always clarifies which is intended. To ensure clarity in comparison analysis, $S_{\mathrm{QMDA}}$ is applied for QMDA.
  
  \item The spectral resolution parameter $L$ of QMDA does \emph{not} coincide with the dimension of the truncated subspace: \citet{giannakis2019qmda} projects all operators onto $\Pi_L : L^2(\mu) \to L^2(\mu)$, mapping into the $(2L{+}1)$-dimensional subspace spanned by $\{\phi_{-L},\dots,\phi_L\}$. Throughout the complexity analysis we follow Appendix~B of the paper, which works directly with $L\times L$ matrices once the orthonormal basis has been fixed; the constant factor of $2$ is therefore absorbed and $L$ is treated as the matrix dimension.
\end{enumerate}

The parameter $q$ denotes the number of integration steps between two consecutive observations and is determined by the physical problem, not by the framework. In QMDA \citep[Appendix~B]{giannakis2019qmda}, the Koopman matrix $\boldsymbol{U}^{(q)}$ is constructed for the \emph{single} value of $q$ required by the assimilation. The paper proposes two implementation strategies:

\begin{enumerate}
  \item compute $\boldsymbol{U}^{(q)}$ directly for the operational $q$ (cost $O(NL^2)$, storage $O(L^2)$);
  \item compute $\boldsymbol{U}^{(1)}$ once and apply the matrix power $\left(\boldsymbol{U}^{(1)}\right)^{q}$ on-the-fly (cost $O(NL^2)$ offline + $O(L^3)$ per application, with risk of numerical instability for large $q$ when $\boldsymbol{U}^{(1)}$ has eigenvalues with positive real part).
\end{enumerate}

In the L63 experiment of \citet{giannakis2019qmda}, $\boldsymbol{U}^{(0)},\dots,\boldsymbol{U}^{(100)}$ are explicitly precomputed to produce output at multiple horizons (Figures~5 and~6 of the reference paper), but this is an experimental choice, not a requirement of the base algorithm.

\paragraph{Convention for asymptotic costs}
Throughout, we use standard big-$O$ notation for worst-case asymptotic complexity and draw a sharp distinction between the \textbf{offline phase} (pre-computation, executed once) and the \textbf{online phase} (assimilation cycle, repeated $K$ times at runtime). 

$O(f(n))$ denotes a worst-case asymptotic upper bound. Costs marked \emph{per cycle} refer to a single pass through the assimilation loop; total online cost follows by multiplying by $K$. Optional operations (e.g.\ the Koopman forecast) are costed separately and excluded from the default totals.

%% file: sec_3_framework_DATO.tex
\section{The DATO Framework}
\label{sec:framework_dato}

DATO \citep{conti2025dato} is a Bayesian filtering framework grounded in
\emph{transfer operator} theory, and proceeds in two phases. In the \textbf{offline phase}, finite-dimensional approximations of the Koopman operator $\mathcal{K}_{\Delta t}$ and of its $L^2(A,\mu)$-adjoint, the Perron--Frobenius operator $\mathcal{P}_{\Delta t}$ (both introduced in \cref{sec:introduzione}), are learned from a training dataset via kEDMD. In the \textbf{online phase}, these representations are used to carry out the Bayesian prediction--analysis cycle.

\subsection{Offline phase}
\label{app:dato:offline}

The offline phase consists of three sequential stages: the construction of the kernel Gram matrices; the eigendecomposition of those matrices to obtain the spectral representation of the transfer operators; and the precomputation of all quantities that will be reused at each assimilation cycle.

The central design principle is that the operator matrices are never materialised explicitly — the entire dynamics is encoded in $S$ scalar eigenvalues — and this choice, made offline, is what drives the efficiency of the online cycle.

\subsubsection{Gram matrix construction}

Given a training trajectory (dataset) $\{x_1,\ldots,x_{m+1}\}$ of $m+1$ snapshots, DATO defines the matrices $X = [x_{1}, \dots, x_{m}]$ and $Y = [x_{2}, ..., x_{m+1}]$ and constructs two Gram matrices $G_{XX},G_{XY}\in\mathbb{R}^{m\times m}$:

\begin{align*}
	(G_{XX})_{ij} & = k(x_i,\,x_j) = \exp\!\left(-\tfrac{\|x_i - x_j\|^2}{2\sigma^2}\right), \\
	(G_{XY})_{ij} & = k(x_i,\,x_{j+1}),
\end{align*}

Each kernel evaluation requires computing the squared Euclidean distance $\|x_i - x_j\|^2$ in $\mathbb{R}^{n}$ at cost $O(n)$, and there are $m^{2}$ pairs per matrix. The matrix $G_{YX}$ needed for the Koopman eigenvalue problem is simply $G_{XY}^{\top}$, obtained at no additional cost. The total Gram matrix construction cost is therefore:

\begin{align}
	T_{\mathrm{Gram}} = O(nm^{2}).
	\label{eq:cost_gram}
\end{align}

A structurally important observation is that the size of these matrices is $m \times m$, independent of the state dimension $n$: the quadratic dependence is in the number of training snapshots, not in the physical dimension of the system. This is the property that allows DATO to scale to moderately high-dimensional systems without modifying the core algorithm.

DATO adopts the unnormalised Gaussian (RBF) kernel

\begin{equation}
  k(x,y)=\exp\!\left(-\frac{\|x-y\|^2}{2\sigma^2}\right),
  \label{eq:dato_kernel}
\end{equation}

which is symmetric, positive definite, and induces a universal RKHS $\mathcal{H}\subset L^2(A,\mu)$ on compact sets. The bandwidth parameter $\sigma$ controls the locality of the representation and is typically set using the \emph{median heuristic} on pairwise distances in the training set \citep{flaxman2016bayesian}.

\paragraph{Eigendecomposition}
Finite-dimensional approximations of the Perron---Frobenius $\mathcal{P}_{\Delta t}$ and Koopman $\mathcal{K}_{\Delta t}$ operators are obtained by solving the two $m \times m$ generalised eigenvalue problems:

\begin{equation}
	\begin{matrix}
		(\mathbf{G}_{XX} + m\varepsilon\,\mathbf{I})^{-1} \mathbf{G}_{XY}\,v^P &= \lambda\,v^P, \\
		(\mathbf{G}_{XX} + m\varepsilon\,\mathbf{I})^{-1} \mathbf{G}_{YX}\,v^K &= \lambda\,v^K,
	\end{matrix}
	\label{eq:eigendecomp}
\end{equation}
	
where $\varepsilon > 0$ is a Tikhonov regularisation parameter introduced to stabilise the inversion of $G_{XX}$.

The computational bottleneck is the factorisation of $(G_{XX} + m \varepsilon I)$. Two strategies are available: direct \textit{Cholesky} factorisation, computed once at cost $O(m^{3})$ and reused for both eigenvalue problems and the downstream linear systems, or, when only the $S$ dominant eigenpairs are needed, \textit{Krylov} methods (ARPACK) at cost $O(m^{2}S)$. \citet{conti2025dato} report empirically a scaling of $O(n_{\mathrm{modes}}^{2.3})$, consistent with implicitly restarted Arnoldi methods on this class of problem. The Krylov advantage becomes operationally significant only when $S \ll m$.

The eigendecomposition cost is therefore:

\[
	T_{\mathrm{eigen}}= \begin{cases*}
		O(m^{3}) & \text{(Cholesky)} \\
		O(m^{2}S) & \text{(Krylov)}
	\end{cases*}
\]

\paragraph{Eigenfunction matrix precomputation}
The Perron–--Frobenius eigenfunctions are expressed, via the reproducing property of the RKHS, through kernel expansions \citep[Sec.2b]{conti2025dato} evaluated at the training points:

\[
	\phi^P_j(x) = \sum_{i=1}^{m} u^P_{j,i}\,k(x,\,x_i),
	\qquad
	\mathbf{u}^P_j = \mathbf{G}_{XX}^{-1}\,v^P_j,
\]

This can be done analogously for $\phi^{\mathcal{K}}_j$. These eigenfunctions form the basis for the spectral representation of densities and observables in the assimilation cycle.

The precomputation of $\Phi \in \mathbb{R}^{m \times S}$, where $\Phi_{ij} = \varphi^{P}_{j}(x_i)$, proceeds in two steps: first, solving $S$ linear systems of size $m \times m$ via back-substitution using the Cholesky factorisation already available, at cost $O(m^{2}S)$; then assembling $\Phi = G_{XX} U^{P}$ as a product between two matrices of size $m \times m$ and $m \times S$, again at cost $O(m^{2}S)$.

Additionally, the normal matrix $\Phi^{\top}\Phi \in \mathbb{R}^{S \times S}$ and its Cholesky factorisation are computed and stored at this stage, at cost $O(mS^{2} + S^{3})$ — an investment that reduces the per-cycle cost of posterior projection during the online phase. The total precomputation cost is:

\[
	T_{\Phi}= O(m^{2}S)
\]

The dominant offline term is therefore $O(m^{3})$ under direct factorisation, or $O(m^{2}S)$ with Krylov methods when $S \ll m$, giving a total offline cost of:

\[
	T_{\mathrm{offline}}^{\mathrm{DATO}}= O(nm^{2} + m^{3})
\]

The $O(nm^{2})$ term for the Gram matrices would only become relevant in extremely high-dimensional systems where $n \gg m$. On the other hand, when $n\ll m$, as in the L63 scenario, the $O(m^3)$ Cholesky factorisation dominates. Table \ref{tab:dato_offline} summarizes the costs.

\begin{table}[htp!]
	\centering
	\renewcommand{\arraystretch}{1.2}\small
	\begin{tabularx}{\columnwidth}{@{}ll@{}}
		\toprule
		\textbf{Operation} & \textbf{Complexity} \\
		\midrule
		Gram matrices & $O(n\cdot m^2)$ \\
		Cholesky factorisation & $O(m^3)$ \\
		Eigendecomposition & $O(m^2 S)$ \\
		Computation of $U^P = \mathbf{G}_{XX}^{-1}V^P$ & $O(m^2 S)$ \\
		Assembly of $\boldsymbol{\Phi} = \mathbf{G}_{XX}U^P$ & $O(m^2 S)$ \\
		\midrule
		\textbf{Total offline} & $O(n\cdot m^2 + m^3)$ \\
		\bottomrule
	\end{tabularx}
	\caption{Computational complexity of the DATO offline phase. $U^{P}$ and $V^{P}$ are respectively two matrices whose columns are the eigenvectors of equations \cref{eq:eigendecomp}.}
	\label{tab:dato_offline}
\end{table}

\subsection{Online phase}
\label{sec:dato_online_sintesi}

Given $S\le m$ dominant eigenpairs $\{\lambda_j,\phi^{\mathcal{P}}_j\}_{j=1}^S$, every sufficiently regular density $\rho(x)\in L^2(A,\mu)$ admits the representation

\begin{equation}
  \rho(x)\approx\sum_{j=1}^S \xi_j\,\phi^{\mathcal{P}}_j(x),
  \label{eq:dato_expansion}
\end{equation}

with $\boldsymbol{\xi}=(\xi_1,\ldots,\xi_S)^\top$ the vector of PF coefficients. 

Each assimilation cycle runs three steps: \emph{prediction} (Perron--Frobenius evolution), \emph{analysis} (Bayesian update), and, optionally, a \emph{Koopman forecast}.

\begin{itemize}
  \item \textbf{Analysis.} Pointwise Gaussian likelihood on $m$ points ($O(m(pn+p^2))$), pointwise update ($O(m)$), projection of the posterior onto the PF basis ($O(mS+S^2)$), state reconstruction ($O(mn)$).
  \item \textbf{Koopman forecast} (optional): $O(mn+mS)$.
\end{itemize}

The total cost per cycle is

\[
  T^{\mathrm{DATO}}_{\mathrm{online}} = O\!\left(mn + mpn + mS + S^2\right).
\]

Which term dominates depends on the regime of $n$, $p$, and $S$; for L63, where $S\sim m$, the posterior projection $O(mS+S^2)$ is the bottleneck. Per-cycle summary: \cref{tab:dato_online}.

\subsubsection{Prediction (Perron--Frobenius evolution)}

The prior density at the next observation time is propagated through the spectral representation of the Perron--Frobenius operator, which acts diagonally on the PF coefficients (see \cref{eq:pf}) \citep[eq.~11]{conti2025dato}.

The powers $\lambda_j^{q}$ depend only on the operator spectrum and on the inter-observation interval $q$; they are computed once offline, leaving the online update to consist of $S$ scalar multiplications. The prior density on the training set is then recovered by a single matrix--vector product against the precomputed eigenfunction matrix $ \rho^b_{k+1} = \boldsymbol{\Phi}\,\xi^{(k+1)} \in \mathbb{R}^m $.

In practice, the density is evaluated only at the training points, so the prior is represented by a vector in $\mathbb{R}^m$.

Therefore, the total cost of the prediction step is

\begin{align}
  T_{\mathrm{pred}} = O(S) + O(mS) = O(mS).
  \label{eq:cost_pred}
\end{align}

The dynamics over the assimilation window is encoded entirely in the $S$ scalars $\{\lambda_j^{q}\}$: the coefficient update costs $O(S)$, while the $O(mS)$ contribution arises solely from the reconstruction of the density on the training set --- the minimum achievable for a spectral expansion of size $S$ evaluated at $m$ points.

\subsubsection{Analysis step (Bayesian update)}

When a new observation $y_{k+1}$ becomes available, the prior density is updated through Bayes' theorem. In DATO, this update is carried out at the training points, after which the resulting posterior is mapped back onto the PF eigenbasis to close the cycle. The analysis step decomposes naturally into four operations: likelihood evaluation, pointwise update, projection onto the PF basis, and reconstruction of the analysis state.

\paragraph{Likelihood evaluation.}
Under the standard assumption of Gaussian observation errors with covariance $\mathbf{R} \in \mathbb{R}^{p\times p}$, the likelihood at training point $x_i$ takes the quadratic form \citep[eq. 14]{conti2025dato}:

\begin{align*}
	\ell_i = \exp\left(-\tfrac{1}{2} \left(H[x_i]-y_{k+1}\right)^{\top} \mathbf{R}^{-1} \left(H[x_i]-y_{k+1}\right)\right),
\end{align*}

with $H[\cdot]$ the observation operator and $i = 1, \ldots, m$.

Each evaluation requires the application of $H$, at cost $O(pn)$, followed by the quadratic product with $\mathbf{R}^{-1}$ (precomputed offline), which contributes a further $O(p^2)$. Aggregating over the $m$ training points, the total computational cost is:

\begin{align}
  T_{\mathrm{lik}} = O\!\left(m(pn + p^2)\right).
  \label{eq:cost_lik}
\end{align}

\paragraph{Pointwise product and normalisation.}
The unnormalised posterior on the training set is obtained by elementwise multiplication of the prior and the likelihood, $\rho^a_{k+1,i} \propto \ell_i\,\rho^b_{k+1,i}$, followed by a global normalisation; both operations scale linearly in the number of training points and contribute $O(m)$ to the total cost.

\paragraph{Projection of the posterior onto the PF basis.}
The posterior, available pointwise on the training set, is re-expressed in the PF eigenbasis by solving the least-squares problem \citep[eq. 15]{conti2025dato}

\[
  \xi^a_{k+1} = \operatorname*{arg\,min}_{\xi} \left\|\rho^a_{k+1} - \boldsymbol{\Phi}\,\xi\right\|_2^2,
\]

whose normal equations involve the $S\times S$ matrix $\boldsymbol{\Phi}^\top\boldsymbol{\Phi}$. Because both $\boldsymbol{\Phi}^\top\boldsymbol{\Phi}$ and its Cholesky factorization are precomputed offline, the online cost reduces to the matrix--vector product $\boldsymbol{\Phi}^\top\rho^a$, which is $O(mS)$, and the back-substitution against the precomputed factor, which is $O(S^2)$:

\begin{align}
  T_{\mathrm{proj}} = O(mS + S^2).
  \label{eq:cost_proj}
\end{align}

Note that the $S^2$ contribution is not dominated by $mS$ and cannot be discarded a priori: whenever $S$ is comparable to $m$ --- as in the L63 configuration of \cref{sec:l63} --- the two terms are of the same order of magnitude and $O(mS + S^2)$, rather than the likelihood term, governs the online complexity of the analysis step (see \cref{eq:cost_analisi_tot}).

\paragraph{Analysis state reconstruction.}
Following \citet{conti2025dato}, the analysis state is defined as the posterior mean evaluated against the training matrix $\mathbf{X} = [x_1, \ldots, x_m]^\top$ \citep[eq. 17]{conti2025dato}:

\[
  x^a_{k+1} = \mathbf{X}^\top\rho^a_{k+1} \in \mathbb{R}^n,
\]

a single matrix--vector product whose cost scales as

\begin{align}
  T_{\mathrm{recon}} = O(mn).
  \label{eq:cost_recon}
\end{align}

\paragraph{Total cost of the analysis step.}
Summing the four contributions, the analysis step has total complexity

\begin{align}
  T_{\mathrm{analisi}} = O\left(mn + mpn + mS + S^2\right).
  \label{eq:cost_analisi_tot}
\end{align}

Which term dominates depends on the relative magnitudes of $S$, $m$, and the observation-related dimensions: in the regime $S \ll m$ and $p \ll n$, the likelihood term $O(mpn)$ is the bottleneck; for $S$ comparable to $m$, as in the L63 experiment, the projection cost $O(mS + S^2)$ takes over.

\subsubsection{Koopman forecast step (optional)}
\label{app:dato:koopman_forecast}

Once the analysis state $x^a_{k+1}$ has been computed, an explicit forecast at the verification time $k+1+\delta$ can be produced directly in observable space. \cite{conti2025dato} note that propagating the state via the Koopman operator yields, in their experiments, more accurate forecasts than propagation via Perron--Frobenius; this section describes the corresponding sequence of operations and quantifies its cost.

The forecast is obtained in four stages, each acting on objects whose dimensions follow naturally from the spectral representation. First, the analysis state is lifted to the RKHS through the kernel vector against the training set,

\[
  \mathbf{k}_*(x^a_{k+1}) = \left[k(x^a_{k+1},x_1),\ldots, k(x^a_{k+1},x_m)\right]^{\top} \in \mathbb{R}^m,
\]

which entails $m$ kernel evaluations, each requiring a Euclidean distance in $\mathbb{R}^n$, for a total cost of $O(mn)$. The Koopman eigenfunctions at the new point are then recovered from the kernel expansion of the RKHS basis, namely

\[
  \phi^K(x^a_{k+1}) = V^K\,\mathbf{k}_*(x^a_{k+1}) \in \mathbb{R}^S,
\]

a single matrix--vector product whose cost is $O(mS)$. Spectral propagation over $\delta$ assimilation windows --- equivalently, $q\cdot\delta$ model steps --- exploits, as in the prediction step, the diagonality of the operator in its own eigenbasis,

\[
  \psi^f = \Lambda^{q\cdot\delta}\,\phi^K(x^a_{k+1}) \in \mathbb{R}^S,
  \qquad \Lambda = \operatorname{diag}(\lambda_1, \ldots,\lambda_S),
\]

and reduces to $S$ scalar multiplications, contributing only $O(S)$ regardless of $n$. The forecast state is finally reconstructed in $\mathbb{R}^n$ as $  x^f = \mathbf{X}^\top\psi^f \in \mathbb{R}^n, $ at a further cost of $O(mn)$.\\

Aggregating the four contributions yields
\begin{align}
	T_{\mathrm{forecast}} = O(mn + mS).
	\label{eq:cost_forecast}
\end{align}

It is worth emphasising that the spectral propagation itself contributes only $O(S)$, and not $O(nS)$: the operator action in the eigenbasis is a pure diagonal rescaling and is independent of the state dimension. The $O(mn)$ component of the total cost is therefore entirely attributable to the kernel evaluation and to the final reconstruction in state space, neither of which involves operator propagation per se.

Combining the contributions of the prediction, analysis, and (optional) Koopman forecast steps yields a per-cycle complexity of

\begin{align}
	T_{\mathrm{online}}^{\mathrm{DATO}} = O\!\left(mn + mpn + mS + S^2\right),
	\label{eq:cost_dato_online}
\end{align}

with the individual operations and their respective costs reported in \cref{tab:dato_online}. The total online cost over $K$ assimilation cycles is obtained, as customary, by linear scaling in $K$.

\begin{table}[htb!]
	\centering
	\begin{tabularx}{\columnwidth}{@{}lll@{}}
		\toprule
		\textbf{Operation} & \textbf{Cost per cycle} \\
		\midrule
		PF prediction         & $O(S)$              \\
		Prior density recons.           & $O(mS)$             \\
		Likelihood on $m$ points              & $O(m(pn+p^2))$      \\
		Bayes. update + norm.        & $O(m)$              \\
		Posterior projection     & $O(mS + S^2)$       \\
		State recons. $x^a$             & $O(mn)$             \\
		\midrule
		Koopman forecast (Opt.)  & $O(mn + mS)$        \\
		\midrule
		\textbf{Total per cycle} & $O(mn + mpn + mS + S^2)$              \\
		\bottomrule
	\end{tabularx}
	\caption{Computational complexity of the DATO online phase (costs per assimilation cycle).}
	\label{tab:dato_online}
\end{table}

Which of the four terms governs the cycle depends on the relative magnitudes of $n$, $p$, $S$, and $m$: the likelihood $O(mpn)$ dominates whenever $pn \gg S$; the posterior projection $O(mS + S^2)$ overtakes when a large fraction of eigenpairs is retained ($S$ comparable to $m$, as in the L63 experiment); and the state reconstruction $O(mn)$ becomes the bottleneck in the low-observation, parsimonious-basis regime ($S \ll m$, $pn \ll m$).

\subsection{Space complexity}
\label{app:dato:memoria}

The memory footprint of DATO is dominated by a small number of structures that persist between offline and online phase, namely the training dataset itself, the kernel Gram matrices, the eigenfunction matrix, and the spectral coefficients carried across assimilation cycles. Their individual sizes follow directly from the dimensional choices already discussed: the training dataset is $m\times n$ and therefore occupies $O(mn)$, each Gram matrix is $m\times m$ and accounts for $O(m^2)$, the eigenfunction matrix $\boldsymbol{\Phi}\in\mathbb{R}^{m\times S}$ contributes $O(mS)$, while the eigenvalues, the PF coefficients and the density evaluated on the training set together require $O(m+S)$ scalars. A complete summary is given in \cref{tab:dato_memoria}.

\begin{table}[htp!]
	\centering
	\begin{tabularx}{\columnwidth}{@{}ll@{}}
		\toprule
		\textbf{Data structure \& Size} & \textbf{Mem. cost} \\
		\midrule
		Training dataset  ($m\times n$)    & $O(mn)$   \\
		Gram matrices  ($m\times m$ each) & $O(m^2)$ \\
		Eigenfunction matrix $\boldsymbol{\Phi}$ ($m\times S$) & $O(mS)$ \\
		Eigenvalues $\{\lambda_j\}$ ($S$ scalars)    & $O(S)$    \\
		PF coefficients $\xi$              ($S$ scalars)    & $O(S)$    \\
		Density $\rho$ on training points  ($m$ scalars)    & $O(m)$    \\
		\midrule
		\textbf{Total}                                   & $O(m^2 + mn)$ \\
		\bottomrule
	\end{tabularx}
	\caption{Space complexity (memory) of DATO.}
	\label{tab:dato_memoria}
\end{table}

Aggregating these contributions, the total memory cost is $O(m^2 + mn)$, with the $O(m^2)$ Gram matrices representing the dominant term whenever $m\gtrsim n$ --- the regime for which DATO is primarily designed. The implications of this scaling are tangible already at moderate training-set sizes: in the L63 configuration adopted by \citet{conti2025dato}, with $m = 2800$, the two Gram matrices alone account for roughly $62\times 10^6$ scalars, corresponding to approximately ${\sim}500$ MB in double precision. This quadratic dependence on $m$, rather than the cost of any single online operation, is what effectively caps the size of the training dataset that can be handled on standard workstations and motivates the use of sparse or low-rank kernel approximations whenever a substantially larger training set is required.

\subsection{Advanced diagnostics}
\label{sec:dato_diag_sintesi}

A distinctive feature of the operator-theoretic formulation is that two diagnostics traditionally available only through ensemble or adjoint computations admit closed-form expressions directly from the analysis density. The first is the \emph{Observation Influence} (OI), $\partial x^a/\partial y = \operatorname{Cov}_{\rho^a}(x, H[x])\,\mathbf{R}^{-1}$, which quantifies the sensitivity of the analysis state to the assimilated observations and which reduces, in the Gaussian-linear limit, to the classical Kalman gain. The second is the \emph{Forecast Sensitivity to Observation Impact} (FSOI), measuring how a perturbation of the analysis-time observations propagates to the verification time through the Koopman dynamics, and corresponds, to first order, to the FSOI conventionally obtained from finite differences between forecast trajectories. Both diagnostics follow directly from the operator structure of the framework and require no auxiliary ensemble integration; we refer to \cite[Sec.~3]{conti2025dato} for the derivations.

\subsection{Assumptions and limitations}
\label{sec:dato_ass}

The validity of DATO hinges on four assumptions, each tied to a specific stage of the construction.

The first is that it requires that the training trajectory adequately samples the region of state space relevant for assimilation; under ergodicity of the underlying dynamics, this is automatically achieved by sampling along a sufficiently long trajectory, but the framework remains applicable whenever training samples are otherwise representative of the invariant measure. The second is the use of a \textbf{universal Gaussian kernel} on a compact domain, which guarantees density of the associated RKHS in $C(A)$ and is what makes the basis expansive enough to represent arbitrary observables and densities. The third is the introduction of a strictly positive \textbf{Tikhonov regulariser} $\varepsilon > 0$, indispensable to stabilise the inversion of the Gram matrix and the resulting generalised eigenvalue problem. The fourth, more subtle but no less important, is the \textbf{representativeness of the training set}: because all densities are evaluated only at the training points, the filtering quality degrades whenever the assimilation trajectory drifts into regions of state space that are poorly covered by the training data.

It is worth emphasising what the framework does \emph{not} require. No assumption is placed on the linearity of the model dynamics, on the Gaussianity of the prior, or on the linearity of the observation operator: the Bayesian update is carried out at the density level and remains valid in fully nonlinear and non-Gaussian regimes, which is precisely the operating regime in which classical Kalman-type filters lose optimality.

%% file: sec_4_framework_QMDA.tex
\section{The QMDA Framework}
\label{sec:framework_qmda}

QMDA (\emph{Quantum Mechanical Data Assimilation}, \cite{giannakis2019qmda}) reformulates sequential data assimilation by transcribing the Dirac--von Neumann axioms of quantum mechanics \citep{takhtajan2008qmm} directly into the setting of a classical, deterministic, partially observed measure-preserving dynamical system. Within this transcription, the deterministic flow between two consecutive measurements is governed by the unitary Koopman operator on $L^2(\mu)$, while the Bayesian update is replaced by the projective dynamics of the von Neumann measurement postulate. The resulting algorithm propagates a trace-class density operator $\hat{\rho}\in\mathbb{R}^{L\times L}$ rather than a coefficient vector, and produces, at each cycle, a discrete probability distribution over a partition of the range of the observed quantity.

The exposition that follows mirrors the structure of \citet{giannakis2019qmda}: \cref{sec:qmda_assiomi} introduces the axiomatic mapping between quantum mechanics and data assimilation; \cref{sec:qmda_quantiz} describes the spectral discretisation that makes the projective update operationally well defined; \cref{sec:qmda_proj} introduces the finite-dimensional projection parametrised by the spectral resolution $L$; \cref{sec:qmda_ddriven} presents the data-driven realisation of the framework, including the construction of the basis from a variable-bandwidth kernel and, when applicable, the use of delay-coordinate maps. Each subsequent subsection corresponds to one stage of the offline or online phase, and reports both the algorithmic content and the associated asymptotic cost. The complete per-stage costs are summarised in \cref{tab:qmda_offline,tab:qmda_online,tab:qmda_memoria}; \cref{sec:qmda_ass} closes the section with a discussion of the structural assumptions and limitations of the framework.

\subsection{From quantum-mechanical axioms to data assimilation axioms}
\label{sec:qmda_assiomi}

Let $\Phi^t : M \to M$ be a continuous, measure-preserving, ergodic flow on a metric space $M$, with an invariant Borel probability measure $\mu$ of compact support. \citet{giannakis2019qmda} maps the canonical axioms QM1--QM5 of Dirac--von Neumann quantum mechanics \cite{takhtajan2008qmm} onto a corresponding set of axioms DA1--DA5 for a \emph{data assimilation system}, as follows.

\begin{description}
  \item[DA1 (Spaces and states).] The reference Hilbert space is $L^2(\mu)$, equipped with the inner product $\langle f, g \rangle_\mu = \int_M f^*\,g \, d\mu$. The states of the system are non-negative trace-class operators $\rho \in B_1(L^2(\mu))$ with $\operatorname{tr}\rho = 1$. The observables are self-adjoint multiplication operators $T_h \in B(L^2(\mu))$ associated with measurement functions $h \in L^\infty(\mu)$ via $T_h f = h\,f$.
  \item[DA2 (Unitary evolution).] Between two measurements, the state evolves under the action of the Koopman unitary group $U^t : L^2(\mu) \to L^2(\mu)$, defined by $U^t f = f \circ \Phi^t$, according to the Heisenberg-picture relation
  \begin{equation}
    \rho_t = U^{t*}\rho_0 U^t.
    \label{eq:qmda_heisenberg}
  \end{equation}
  \item[DA3 (Spectrum and projection-valued measure).] To each observable $A = T_h$ is associated a projection-valued measure $E_h : \mathcal{B}(\mathbb{R}) \to B(L^2(\mu))$, and the spectrum $\sigma(T_h)$ coincides with the essential range of $h$.
  \item[DA4 (Measurement probability).] If the system is in state $\rho \in B_1(L^2(\mu))$, the probability that a measurement of $A$ yields a value in $\Omega \subseteq \mathbb{R}$ is $\operatorname{tr}(E_A(\Omega)\,\rho)$.
  \item[DA5 (Projective dynamics).] If immediately before a measurement the system is in state $\rho^-$, and the measurement of $A$ yields the value $a \in \sigma_p(A)$ with $E_A(\{a\}) \neq 0$, then the post-measurement state is
  \begin{equation}
    \rho^+ = \frac{E_A(\{a\})\,\rho^-\,E_A(\{a\})}{\operatorname{tr}\!\left(E_A(\{a\})\,\rho^-\,E_A(\{a\})\right)}.
    \label{eq:qmda_projective}
  \end{equation}
\end{description}

Axioms DA2 and DA5 encode the unitary and projective parts of the dynamics, respectively, and constitute the QMDA analogues of the classical forecast and analysis steps of sequential data assimilation. The mapping replaces three structural ingredients of the classical formulation: the probabilistic state ceases to be a Borel measure on state space and becomes a density operator on $L^2(\mu)$; the deterministic evolution is realised through the Koopman action rather than as a push-forward of measures; and the Bayesian update is implemented as a trace-renormalised projection rather than as a pointwise product of prior and likelihood.

\subsection{Spectral discretisation through observable quantisation}
\label{sec:qmda_quantiz}

Axiom DA5 is, as stated, applicable only to measurements lying in the point spectrum $\sigma_p(A)$. To handle observables with a continuous spectrum --- the rule rather than the exception in dynamical applications --- \citet{giannakis2019qmda} introduces a conditional averaging procedure that yields a quantised observable with purely point spectrum. Given $h \in L^\infty(\mu)$ and a uniform partition $\{J_0, \ldots, J_{S-1}\}$ of $(0,1)$ into intervals of equal length $1/S$, one induces the partition of $\mathbb{R}$
\begin{equation}
  \Xi_i = \operatorname{cdf}_h^{-1}(J_i), \qquad M_i = h^{-1}(\Xi_i),
\end{equation}
with $\mu(M_i) = 1/S$ by construction. The conditional expectation $\bar h = \mathbb{E}(h \mid \pi) = \sum_{i=0}^{S-1} \bar a_i\,\mathbf{1}_{M_i}$, with cell averages $\bar a_i = \int_{M_i} h\,d\mu$ and partition membership map $\pi : M \to \{0,\ldots,S-1\}$, defines a quantised observable. The associated multiplication operator $T_{\bar h}$ has purely point spectrum and an atomic projection-valued measure
\begin{equation}
  E_{\bar h}(\{\bar a_i\}) = E_{\bar h}(\Xi_i) = T_{\mathbf{1}_{M_i}}.
\end{equation}
Axiom DA5 is then applied to $T_{\bar h}$ in place of $T_h$, yielding the operational version DA5$'$. The measurement probabilities are unaffected by quantisation in the sense that $P_i(t) = \operatorname{tr}(E_h(\Xi_i)\,\rho_t) = \operatorname{tr}(E_{\bar h}(\{\bar a_i\})\,\rho_t)$, so DA4 remains intact.

\subsection{Finite-dimensional projection and spectral resolution $L$}
\label{sec:qmda_proj}

To make the formalism numerically tractable, one fixes a spectral resolution parameter $L \in \mathbb{N}$ and composes every operator with the orthogonal projection $\Pi_L : L^2(\mu) \to L^2(\mu)$ onto the $(2L{+}1)$-dimensional subspace spanned by the first $L$ Koopman eigenfunctions (in the data-driven discrete formulation, the subspace is $L$-dimensional and the constant factor of $2$ is absorbed into the matrix dimension). Under this projection, the projected unitary evolution and projective update of DA2 and DA5$'$ take the form
\begin{align}
  \hat{\rho}_t &= \frac{U_L^{t*}\,\hat{\rho}_0\,U_L^{t}}{\operatorname{tr}\!\left(U_L^{t*}\,\hat{\rho}_0\,U_L^{t}\right)}, \qquad U_L^{t} = \Pi_L\,U^{t}\,\Pi_L, \\
  \hat{P}_i(t) &= \operatorname{tr}\!\left(E_{h,L}(\{a_i\})\,\hat{\rho}_t\right), \qquad E_{h,L}(\Omega) = \Pi_L\,E_h(\Omega)\,\Pi_L, \\
  \hat{\rho}_i^{+} &= \frac{E_{h,L}(\{a_i\})\,\hat{\rho}^{-}\,E_{h,L}(\{a_i\})}{\operatorname{tr}\!\left(E_{h,L}(\{a_i\})\,\hat{\rho}^{-}\,E_{h,L}(\{a_i\})\right)}.
  \label{eq:qmda_update}
\end{align}
A subtle but consequential point, made explicit by \citet[Sec.~IV]{giannakis2019qmda}, concerns the trace renormalisation in the projected evolution: ``\textit{the division by $\operatorname{tr}(U_L^{t*}\rho_0 U_L^t)$ in the expression for $\hat{\rho}_t$ is due to the fact that, unlike $U^t$, $U_L^{t}$ is not unitary, and thus does not preserve the trace of $\rho_0$}''. The renormalisation is therefore not optional but structural: it compensates for the loss of unitarity induced by the truncation. In the limit $L \to \infty$, all expressions converge to their infinite-dimensional counterparts.

\subsection{Data-driven formulation}
\label{sec:qmda_ddriven}

In realistic applications neither the invariant measure $\mu$ nor the equations of motion are accessible in closed form, and $\mu$ is often supported on a fractal attractor of non-integer dimension. QMDA therefore constructs every ingredient of the framework directly from a time-ordered sequence $F(x_0), F(x_1), \ldots, F(x_{N-1})$ of observations along a trajectory $x_n = \Phi^{n\Delta t}(x_0)$, where $F : M \to Y$ is an injective observation map taking values in a data space $Y$, conventionally $Y = \mathbb{R}^d$.

\paragraph{Empirical measure and discrete Hilbert space.}
The trajectory is associated with the sampling measure $\mu_N = \frac{1}{N}\sum_{n=0}^{N-1}\delta_{x_n}$, which by ergodicity converges weakly to $\mu$ as $N \to \infty$. The space $L^2(\mu_N)$ is isometrically isomorphic to $\mathbb{C}^N$ equipped with the inner product $\langle \vec{f}, \vec{g} \rangle = \vec{f}^*\vec{g}/N$, and every linear operator on $L^2(\mu_N)$ admits a representation as an $N \times N$ matrix.

\paragraph{Variable-bandwidth kernel and bistochastic normalisation.}
The data-driven basis is constructed from a kernel $p_N : M \times M \to \mathbb{R}$ defined as a pullback from the data space, $p_N(x, x') = \tilde p_N(F(x), F(x'))$, that is required to be $L^2(\mu_N)$-strictly positive and Markov ergodic. QMDA starts from the variable-bandwidth Gaussian kernel of \citet{berry2016vbkernel},
\begin{equation}
  \tilde k_N(y, y') = \exp\!\left(-\frac{d^{\,2}(y, y')}{\epsilon\,\sigma_N(y)\,\sigma_N(y')}\right),
  \label{eq:qmda_vbkernel}
\end{equation}
with $d$ the Euclidean distance on $Y$ and $\sigma_N : Y \to \mathbb{R}_{+}$ a continuous, positive bandwidth function adapted to the local sampling density. The unnormalised kernel is then rendered Markovian through the symmetric bistochastic normalisation of \citet{coifman2013bistochastic}. The eigendecomposition of the integral operator $G_{\mu_N}$ associated with $p_N$ provides an orthonormal basis $\{\phi_{j,N}\}_{j=0}^{N-1}$ of $L^2(\mu_N)$ whose continuous representatives converge, in the limit $N \to \infty$, to a basis of $L^2(\mu)$ qualitatively analogous to that of the Laplace--Beltrami eigenfunctions, in the sense of pointwise spectral convergence \cite[Theorem~1, Appendix~A]{giannakis2019qmda}.

\paragraph{Data-driven operators.}
Within the data-driven framework, the continuous operators of \cref{sec:qmda_proj} are replaced by their empirical counterparts. The unitary group $U^t$ at lag $t = q\Delta t$ becomes the $q$-step shift operator $U_N^{(q)} : L^2(\mu_N) \to L^2(\mu_N)$, defined by $(U_N^{(q)} f)(x_n) = f(x_{n+q})$ for $0 \le n \le N - q - 1$ and zero otherwise. The cumulative distribution of $h$ is replaced by its empirical analogue $\operatorname{cdf}_{h,N}$, which induces the empirical partition $\Xi_{i,N}$ and the corresponding quantisation $\bar h_N$. After fixing $L \le N - 1$ and the projection $\Pi_{L,N} : L^2(\mu_N) \to \operatorname{span}(\phi_{0,N}, \ldots, \phi_{L-1,N})$, one defines the projected operators $U_{L,N}^{(q)} = \Pi_{L,N}\,U_N^{(q)}\,\Pi_{L,N}$ and $E_{\bar h_N, L} = \Pi_{L,N}\,E_{\bar h_N}\,\Pi_{L,N}$, which together yield a data-driven realisation of DA2 and DA5$'$ with trace renormalisation identical in form to \eqref{eq:qmda_update}.

\paragraph{Delay-coordinate maps.}
When an injective observation map $F$ is unavailable and only a scalar measurement function $h : M \to \mathbb{R}$ is at hand, QMDA recovers injectivity through delay-coordinate embedding \cite{takens1981delay,sauer1991embedology,robinson2005embedding}: for $Q \in \mathbb{N}$,
\begin{equation}
  h_Q(x) = \bigl(h(x),\,h(\Phi^{-\Delta t}(x)),\,\ldots,\,h(\Phi^{-(Q-1)\Delta t}(x))\bigr) \in \mathbb{R}^Q.
  \label{eq:qmda_delay}
\end{equation}
Under mild hypotheses on $\Phi^t$, $h$, and $\Delta t$, there exists $Q_*$ such that $h_Q$ is injective on every compact set whenever $Q > Q_*$. QMDA then sets $F = h_Q$ with $Q$ chosen sufficiently large, with the practical implication that the data-space dimension $d$ is replaced by $Q$ in every offline cost expression.

\subsection{Convergence of the data-driven scheme}
\label{sec:qmda_conv}

Theorem~1 of \citet[Appendix~A]{giannakis2019qmda} establishes that, under axioms DA1--DA3 and for kernels $p_N$ satisfying the structural properties listed above, the matrix elements of $U_{L,N}^{(q)}$ and $E_{\bar h_N, L}$, the partition intervals $\Xi_{i,N}$, and the membership assignments $\pi_N(a)$ all converge to their continuous counterparts in the limit $N \to \infty$ at fixed $L$, and to the full infinite-dimensional scheme in the iterated limit $N \to \infty$ followed by $L \to \infty$. Since convergence is non-uniform in $j, k$, $L$ must be chosen substantially smaller than $N$ in practice ($L/N \sim 10^{-2}$ in the L63 experiment of \citet{giannakis2019qmda}).

\subsection{Offline phase}
\label{sec:qmda_offline}

The offline phase of QMDA proceeds through three logically distinct stages: the construction of a data-driven kernel matrix from the observations, the eigendecomposition of that matrix to obtain a basis for $L^{2}(\mu_{N})$, and the explicit materialisation of the operator matrices that drive the online cycle. The third stage is the architecturally decisive one: by precomputing the Koopman matrices and the spectral projectors once and for all, QMDA decouples the online cost from the training-set size $N$ and from the physical state dimension $n$, which is the structural property responsible for the favourable scaling behaviour of the framework at runtime.

\subsubsection{Kernel construction}
\label{sec:qmda_kernel}

Starting from the time-ordered observation sequence $\{F(x_n)\}_{n=0}^{N-1} \subset \mathbb{R}^d$, QMDA assembles the $N \times N$ kernel matrix $\mathbf{G}$ associated with the variable-bandwidth Gaussian kernel \eqref{eq:qmda_vbkernel}, followed by the symmetric bistochastic (Sinkhorn) normalisation. The brute-force assembly of $\mathbf{G}$ requires the evaluation of $N^2$ kernel entries, each involving a Euclidean distance in $\mathbb{R}^d$, which gives a cost of $O(d\,N^2)$, as reported explicitly in \cite[Appendix~B]{giannakis2019qmda}.

The nominal $O(N^{2})$ memory footprint of the dense kernel matrix would be prohibitive at the operational training-set sizes adopted by \citet{giannakis2019qmda} --- at $N = 64\,000$, a dense $\mathbf{G}$ would already exceed $32$~GB in double precision --- and is therefore replaced by a sparse approximation $\hat{\mathbf{G}}$ that retains, for each row, only the $r$ nearest neighbours of the corresponding point. The approximation rests on the rapid decay of the Gaussian kernel and reduces the storage requirement to $O(rN)$; in the L63 experiment of \citet{giannakis2019qmda}, $r = 5\,000 \approx 8\%$ of $N$. The bistochastic normalisation is then carried out by an iterative Sinkhorn procedure operating on $\hat{\mathbf{G}}$, at a cost of $O(rN k_{\mathrm{iter}})$ that is typically negligible compared with the kernel-matrix assembly.

The dominant cost of this stage is therefore
\begin{align}
  T_{\mathrm{kernel}}^{\mathrm{QMDA}} = O(d\,N^{2}).
  \label{eq:cost_kernel_qmda}
\end{align}

\subsubsection{Eigendecomposition}
\label{sec:qmda_eigen}

The leading $L$ eigenvectors of the sparse kernel matrix $\hat{\mathbf{G}}$ are computed via the implicitly restarted Arnoldi method, as implemented in ARPACK \cite{lehoucq1998arpack}. The sparsity of $\hat{\mathbf{G}}$ reduces the cost of every matrix--vector product from $O(N^{2})$ to $O(rN)$, and convergence to the leading $L$ eigenpairs requires $O(L)$ Arnoldi iterations, with the constant depending on the spectral gap. Aggregating these contributions yields
\begin{align}
  T_{\mathrm{eigen}}^{\mathrm{QMDA}} = O(L\,r\,N).
  \label{eq:cost_eigen_qmda}
\end{align}
The resulting eigenvectors $\{\tilde{\phi}_{j,N}\}_{j=0}^{L-1}$ form a data-driven orthonormal basis of $L^{2}(\mu_{N})$ whose continuous representatives, in the limit $N \to \infty$, approach a Laplace--Beltrami-like basis of $L^{2}(\mu)$ \cite[Theorem~1]{giannakis2019qmda}.

\subsubsection{Operator materialisation}
\label{sec:qmda_operators}

Once the basis is available, QMDA constructs explicitly all $L \times L$ operator matrices that will be invoked at runtime. For the Koopman operator at the assimilation lag $q$, fixed by the inter-observation interval of the physical problem, one forms \cite[Appendix~B, S2]{giannakis2019qmda}
\[
  U^{(q)}_{jk} = \frac{1}{N}\sum_{n=0}^{N-q-1} \vec{\phi}_{j,n}\,\vec{\phi}_{k,n+q}, \qquad 0 \le j, k \le L - 1,
\]
which evaluates to $O(NL^{2})$ operations: each of the $L^{2}$ entries demands a sum over $N - q$ terms. \citet{giannakis2019qmda} discusses two implementation strategies for handling the temporal evolution: either the matrix $U^{(q)}$ is computed directly for the single operational value of $q$ --- the strategy adopted as the asymptotic baseline of this analysis --- or the matrix $U^{(1)}$ is computed once and the matrix power $(U^{(1)})^{q}$ is applied on the fly, at an additional cost of $O(L^{3})$ per application and at the price of potential numerical instability for large $q$ when $U^{(1)}$ has eigenvalues with positive real part. In the L63 experiment, the matrices $U^{(0)}, \ldots, U^{(100)}$ are precomputed explicitly to produce output at multiple horizons, raising the operator-construction cost to $101 \times O(NL^{2})$; this is, however, an experimental choice rather than a structural requirement, and the asymptotic baseline retains
\begin{align}
  T_{U^{(q)}}^{\mathrm{QMDA}} = O(NL^{2}).
  \label{eq:cost_Uq_qmda}
\end{align}

For each of the $S_{\mathrm{QMDA}}$ elements $\Xi_i$ of the observation partition, the spectral projector $\mathbf{E}_i \in \mathbb{R}^{L \times L}$ is given by \cite[Appendix~B, S2]{giannakis2019qmda}
\[
  \mathbf{E}_{i,jk} = \frac{1}{N}\sum_{n \in N_i}\vec{\phi}_{j,n}\,\vec{\phi}_{k,n}, \qquad N_i = \{n : h(x_n) \in \Xi_i\}.
\]
A single projector $\mathbf{E}_i$ is assembled at cost $|N_i| \cdot L^{2}$, corresponding to a sum of $|N_i|$ outer products of size $L \times L$. Since the index sets $\{N_i\}_{i=1}^{S_{\mathrm{QMDA}}}$ partition $\{0, \ldots, N-1\}$ and therefore $\sum_i |N_i| = N$, the aggregate cost over all projectors is
\begin{align}
  T_{\mathbf{E}}^{\mathrm{QMDA}} = \sum_{i=1}^{S_{\mathrm{QMDA}}} |N_i| \cdot L^{2} = N \cdot L^{2} = O(NL^{2}),
  \label{eq:cost_proj_qmda}
\end{align}
in agreement with \cite[Appendix~B, S2]{giannakis2019qmda}.

\subsubsection{Offline phase summary}
\label{sec:qmda_offline_summary}

Aggregating the contributions of \eqref{eq:cost_kernel_qmda}, \eqref{eq:cost_eigen_qmda}, \eqref{eq:cost_Uq_qmda} and \eqref{eq:cost_proj_qmda} yields the total offline complexity
\[
  T_{\mathrm{offline}}^{\mathrm{QMDA}} = O(d\,N^{2} + N\,L^{2}),
\]
in which the kernel-matrix assembly $O(d\,N^{2})$ dominates whenever $N \gg L$ --- the operationally relevant regime of the framework. The individual contributions are summarised in \cref{tab:qmda_offline}.

\begin{table}[htp!]
  \centering
  \begin{tabularx}{\columnwidth}{@{}lX@{}}
    \toprule
    \textbf{Operation} & \textbf{Complexity} \\
    \midrule
    Kernel matrix $\mathbf{G}$             & $O(d\,N^{2})$ \\
    Bistochastic normalisation                 & $O(r\,N\,k_{\mathrm{iter}})$ \\
    Sparse eigenvectors              & $O(L\,r\,N)$ \\
    Koopman matrix $U^{(q)}$     & $O(N\,L^{2})$ \\
    Spectral projectors $\{\mathbf{E}_i\}_{i=1}^{S_{\mathrm{QMDA}}}$  & $O(N\,L^{2})$ \\
    \midrule
    \textbf{Total offline}                   & $O(d\,N^{2} + N\,L^{2})$ \\
    \bottomrule
  \end{tabularx}
  \caption{Computational complexity of the QMDA offline phase. The Koopman cost refers to constructing a single $U^{(q)}$.}
  \label{tab:qmda_offline}
\end{table}

The structural implication of this decomposition is that all dependence on the training-set size $N$, on the data-space dimension $d$, and ultimately on the physical state dimension $n$ is absorbed offline. Once $U^{(q)}$ and the projectors $\{\mathbf{E}_i\}$ are stored, the online cycle operates entirely in the $L$-dimensional spectral subspace, and its cost is governed by $L$ alone, as the next subsection shows.

\subsection{Online phase}
\label{sec:qmda_online}

In contrast with DATO, which carries a coefficient vector $\xi \in \mathbb{R}^{S}$, QMDA propagates a full density operator $\hat{\rho} \in \mathbb{R}^{L \times L}$ from one assimilation cycle to the next. This structural difference dictates the entire complexity profile of the online phase: each cycle reduces to a small number of operations on $L \times L$ matrices, namely the unitary evolution between observations, the evaluation of the measurement probabilities, and the projective update upon arrival of a new observation.

\subsubsection{Density-operator evolution}
\label{sec:qmda_evolution}

Between two successive observations, the density operator evolves under the Schrödinger-picture dynamics \citep[Sec. V and Appendix B, S3]{giannakis2019qmda}
\[
  \hat{\rho} = \frac{U_{L}^{(q)*}\,\hat{\rho}^{+}\,U_{L}^{(q)}}{\operatorname{tr}\!\left(U_{L}^{(q)*}\,\hat{\rho}^{+}\,U_{L}^{(q)}\right)}.
\]
The trace renormalisation is required because the projected operator $U_L^{(q)}$ is the rank-$L$ truncation of a unitary operator on $L^2(\mu)$ and is therefore not itself unitary, as \citet[Sec.~IV]{giannakis2019qmda} makes explicit. Computationally, the dominant contribution comes from two consecutive $L \times L$ matrix--matrix products, and the trace evaluation is a lower-order term, giving
\begin{align}
  T_{\mathrm{ev}}^{\mathrm{QMDA}} = O(L^{3}).
  \label{eq:cost_ev_qmda}
\end{align}

\subsubsection{Measurement probabilities}
\label{sec:qmda_prob}

The probability that the observable $h$ falls within the partition element $\Xi_i$ at the current time follows from axiom DA4 specialised to the projected operators \cite[eq.~4 and Appendix~B, S3]{giannakis2019qmda}:
\[
  \hat{P}_{i, N}(t) = \operatorname{tr}\!\left(\mathbf{E}_i\,\hat{\rho}\right) = \sum_{j, k} (\mathbf{E}_i)_{jk}\,\hat{\rho}_{kj}.
\]
\citet[Appendix~B, S3]{giannakis2019qmda} states the aggregate cost over the $S_{\mathrm{QMDA}}$ partition elements as
\begin{align}
  T_{\mathrm{prob}}^{\mathrm{QMDA}} = O(S_{\mathrm{QMDA}}\,L),
  \label{eq:cost_prob_qmda}
\end{align}
without explicit derivation; the figure is consistent with the dominant contribution coming from $S_{\mathrm{QMDA}}$ trace evaluations, each requiring $O(L)$ operations when the sparsity structure of the projectors $\mathbf{E}_i$ is exploited.

\subsubsection{Projective update}
\label{sec:qmda_update}

When the measurement of $h$ at time $t_{n+1}$ is found to lie in $\Xi_i$, the density operator is updated via the projective dynamics of axiom DA5$'$ \cite[Appendix~B, S3]{giannakis2019qmda},
\[
  \hat{\rho}^{+} = \frac{\mathbf{E}_i\,\hat{\rho}^{-}\,\mathbf{E}_i}{\operatorname{tr}\!\left(\mathbf{E}_i\,\hat{\rho}^{-}\,\mathbf{E}_i\right)},
\]
whose computational cost is once again dominated by two $L \times L$ matrix--matrix products and is therefore
\begin{align}
  T_{\mathrm{upd}}^{\mathrm{QMDA}} = O(L^{3}).
  \label{eq:cost_upd_qmda}
\end{align}

\subsubsection{Online phase summary}
\label{sec:qmda_online_summary}

Combining \eqref{eq:cost_ev_qmda}, \eqref{eq:cost_prob_qmda} and \eqref{eq:cost_upd_qmda} yields the per-cycle complexity
\[
  T_{\mathrm{online}}^{\mathrm{QMDA}} = O(L^{3}), \qquad \text{for } L \gg S_{\mathrm{QMDA}},
\]
and the total online cost over $K$ assimilation cycles scales linearly in $K$. The individual contributions are reported in \cref{tab:qmda_online}.

\begin{table}[htp!]
  \centering
  \renewcommand{\arraystretch}{1.25}\small
  \begin{tabularx}{\columnwidth}{@{}lX@{}}
    \toprule
    \textbf{Operation} & \textbf{Cost per cycle} \\
    \midrule
    Unitary evolution (Schrödinger picture)              & $O(L^{3})$ \\
    Measurement probabilities ($S_{\mathrm{QMDA}}$ bins) & $O(S_{\mathrm{QMDA}}\,L)$ \\
    Projective update (analysis)                         & $O(L^{3})$ \\
    \midrule
    \textbf{Total per cycle} ($L \gg S_{\mathrm{QMDA}}$) & $O(L^{3})$ \\
    \textbf{Total online} ($K$ cycles)                   & $K\cdot O(L^{3})$ \\
    \bottomrule
  \end{tabularx}
  \caption{Computational complexity of the QMDA online phase (costs per assimilation cycle). The measurement-probability cost is the conservative estimate derivable from the definition of the $\mathbf{E}_i$ entries; \citet{giannakis2019qmda} reports $O(S_{\mathrm{QMDA}}\,L)$ without explicit derivation.}
  \label{tab:qmda_online}
\end{table}

This per-cycle cost is independent of the physical state dimension $n$, of the training-set size $N$, of the data-space dimension $d$, and of the assimilation lag $q$. The dimension-independence of the online cycle is the most operationally significant property of the framework: once the offline precomputation has been completed, the cost of the assimilation loop is pinned to the spectral resolution $L$, irrespective of how large the physical state space becomes. This is precisely the regime in which QMDA is structurally advantageous compared with classical state-space filters, whose per-cycle cost scales explicitly with $n$.

\subsection{Space complexity}
\label{sec:qmda_memoria}

The memory footprint of QMDA is concentrated on a small number of structures that persist across the offline and online phases: the sparse kernel matrix used during basis construction, the eigenvector matrix providing the orthonormal basis of $L^{2}(\mu_{N})$, the density operator that constitutes the runtime state, and the precomputed Koopman matrix and spectral projectors. The dominant contributions come from the eigenvector matrix, which is $N \times L$ and therefore occupies $O(NL)$, and from the collection of $S_{\mathrm{QMDA}}$ projectors, each $L \times L$, which together account for $O(S_{\mathrm{QMDA}}\,L^{2})$. The remaining structures --- the sparse kernel matrix at $O(rN)$ and the $L \times L$ operators at $O(L^{2})$ each --- are subdominant in the operational regime of the framework. Table \ref{tab:qmda_memoria} reports the complete breakdown.

\begin{table}[htp!]
  \centering
  \renewcommand{\arraystretch}{1.2}
  \begin{tabularx}{\columnwidth}{@{}XX@{}}
    \toprule
    \textbf{Data structure} & \textbf{Memory cost} \\
    \midrule
    Sparse kernel matrix $\hat{\mathbf{G}}$      & $O(rN)$ \\
    Eigenvector matrix $\tilde{\phi}_j$            & $O(NL)$ \\
    Density operator $\hat{\rho}$                & $O(L^{2})$ \\
    Koopman matrix $U^{(q)}$                & $O(L^{2})$ \\
    Spectral projectors $\{\mathbf{E}_i\}$    & $O(S_{\mathrm{QMDA}}\,L^{2})$ \\
    \midrule
    \textbf{Total}                                               & $O(NL + S_{\mathrm{QMDA}}\,L^{2})$ \\
    \bottomrule
  \end{tabularx}
  \caption{Space complexity (memory) of QMDA. Practical values for the L63 experiment: $N = 64\,000$, $L = 1\,000$, $S_{\mathrm{QMDA}} = 32$, single $U^{(q)}$.}
  \label{tab:qmda_memoria}
\end{table}

The total memory cost is therefore $O(NL + S_{\mathrm{QMDA}}\,L^{2})$, a profile qualitatively distinct from the $O(m^{2} + mn)$ scaling of DATO: where DATO is bottlenecked by the quadratic dependence on the training-set size induced by the Gram matrices, QMDA is bottlenecked by the bilinear $NL$ term from the eigenvector matrix, with the projectors providing a subdominant $L^{2}$ contribution scaled by $S_{\mathrm{QMDA}}$. The single Koopman matrix occupies only $O(L^{2})$ and is negligible against the basis-related terms, although the precomputation of multi-horizon matrices $U^{(0)}, \ldots, U^{(q)}$ raises this contribution to $|\mathcal{Q}|\,L^{2}$. Concrete byte-level estimates for the L63 configuration are reported in \cref{sec:l63}.

When only a scalar observable $h : M \to \mathbb{R}$ is available, the use of delay-coordinate maps replaces the data-space dimension $d$ with the number of delays $Q$ in every offline cost expression, so that the kernel construction cost becomes $O(Q\,N^{2})$. The memory profile is otherwise unaffected.

\subsection{Assumptions and limitations}
\label{sec:qmda_ass}

The validity of QMDA rests on a sequence of assumptions, each tied to a specific stage of the construction. The first is the \textbf{ergodicity and measure preservation} of the underlying flow $\Phi^t$ with respect to the invariant measure $\mu$, which is what makes the empirical sampling measure $\mu_N$ converge weakly to $\mu$ and what underpins every spectral statement of the framework. Closely related is the requirement that $\mu$ be a \textbf{Borel probability measure with compact support}: crucially, $\mu$ need not be absolutely continuous with respect to any reference measure, so QMDA accommodates measures supported on fractal attractors --- the L63 case being a paradigmatic example --- without modification of the algorithm.

The third assumption concerns the \textbf{injectivity of the observation map} $F$, necessary for the completeness of the data-driven basis; when only a scalar observable $h$ is at hand, injectivity is recovered through delay-coordinate maps with a sufficient number of delays $Q$, at the price of introducing $Q$ as an effective data-space dimension. It must be noted that the observation error covariance matrix $R$ is not used.

The fourth concerns the \textbf{kernel construction}: the kernel must be $L^2$-strictly positive and Markov ergodic, conditions met by the variable-bandwidth Berry--Harlim kernel \cite{berry2016vbkernel} after symmetric bistochastic normalisation \cite{coifman2013bistochastic}. Finally, convergence to the infinite-dimensional scheme requires the double limit $N \to \infty$ followed by $L \to \infty$ (\cref{sec:qmda_conv}).

Notably absent from this list are any Gaussianity or linearity hypotheses on the dynamics, on the observation operator, or on the prior or posterior distributions: in perfect analogy with the quantum-mechanical formalism, QMDA is intrinsically free from such restrictions, which is what makes it a natural candidate for non-Gaussian and strongly nonlinear regimes where Kalman-type filters lose optimality.

%% file: sec_5_comparison.tex
\section{Comparison between DATO and QMDA}
\label{sec:confronto}

\begin{figure*}[htb!]
  \centering
  \includegraphics[width=0.65\textwidth]{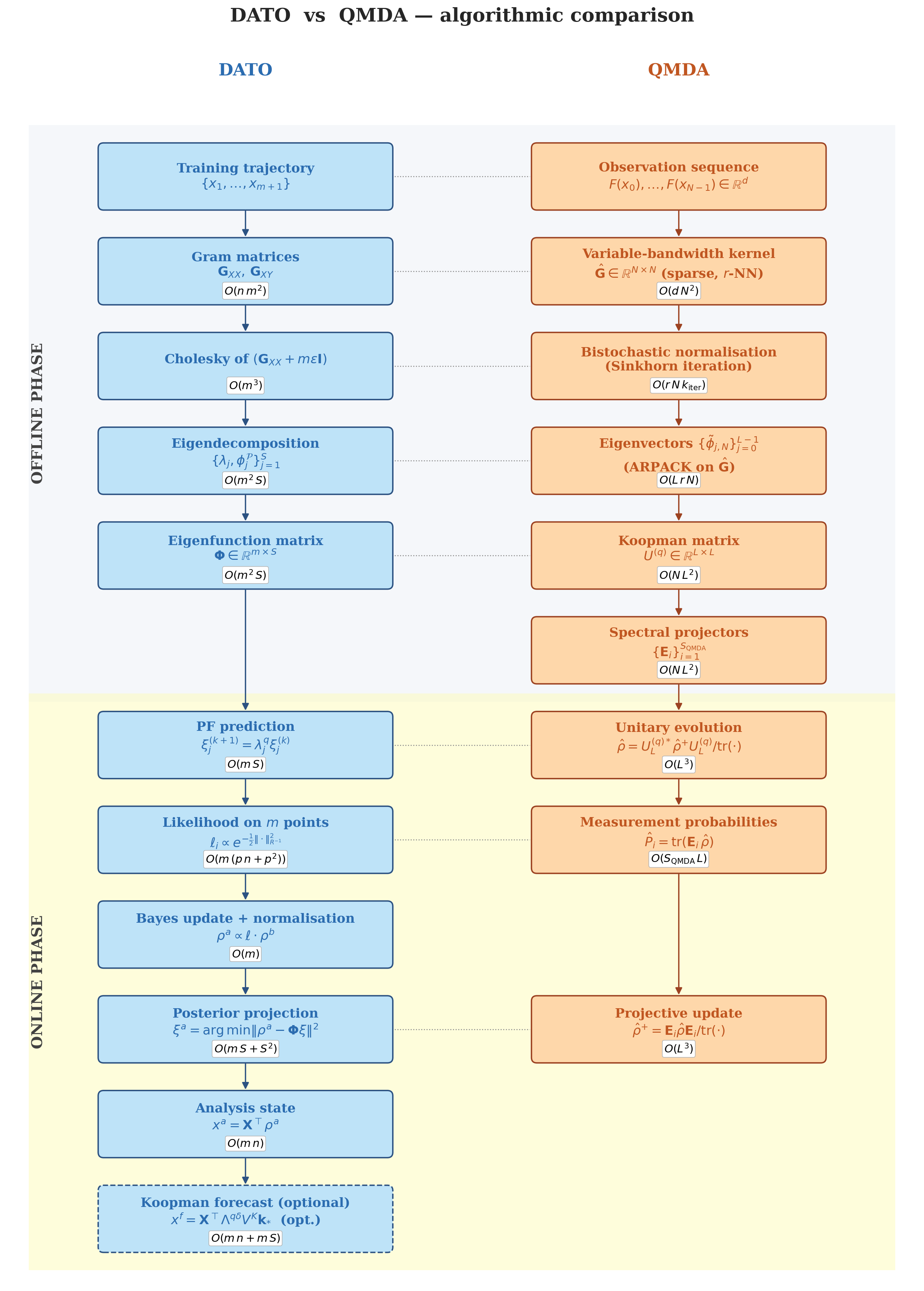}
  \caption{Side-by-side comparison of the DATO (left) and QMDA (right) algorithmic pipelines. The two background bands separate the offline phase (top) from the online cycle (bottom). Each node reports the operation, its formal expression, and the asymptotic cost; dotted horizontal connectors link conceptually analogous steps across the two frameworks. The structural mismatches are also visible at a glance: QMDA introduces an additional offline stage for the materialisation of the spectral projectors $\{\mathbf{E}_i\}$, while DATO carries extra online operations --- the analysis-state reconstruction and the optional Koopman forecast (dashed box) --- that have no direct counterpart in QMDA.}
  \label{fig:flowchart_comparison}
\end{figure*}

Despite sharing a common root in Koopman operator theory, DATO and QMDA differ profoundly by design in how the system state is represented, how the Bayesian update mechanism is implemented, and what the output of each assimilation cycle is. These differences are not merely notational: they determine the entire computational profile of each framework. The main conceptual correspondences are summarised in \cref{tab:corrispondenze}, while \cref{fig:flowchart_comparison} provides a side-by-side visual map of the two algorithmic pipelines.

\begin{table}[htp!]
	\centering
	\small
	\begin{tabularx}{\columnwidth}{@{}lXX@{}}
		\toprule
		& \textbf{Concept} \\
		\midrule
		& \textbf{Hilbert space} \\
		DATO & Gaussian kernel RKHS \\
		QMDA & Normalised $L^2(\mu_N)$ \\
		\midrule
		
		& \textbf{Evolution operator} \\
		DATO & PF + Koopman (separate) \\
		QMDA & Unitary Koopman $U^t$ \\
		\midrule
		
		& \textbf{Finite-dim. approximation} \\
		DATO & Implicit kEDMD, $O(m^2)$ \\
		QMDA & Sparse eigenvectors, $O(LrN)$ \\
		\midrule
		
		& \textbf{System state} \\
		DATO & PF coeff.\ $\xi\in\mathbb{R}^S$ \\
		QMDA & Density op.\ $\hat{\rho}\in\mathbb{R}^{L\times L}$ \\
		\midrule
		
		& \textbf{Update} \\
		DATO & Likelihood $\times$ prior \\
		QMDA & Projective dynamics \\
		\midrule
		
		& \textbf{Discretisation} \\
		DATO & Gaussian likelihood \\
		QMDA & Partition into bins \\
		\midrule
		
		& \textbf{Output} \\
		DATO & Estimate $x^a$ + density \\
		QMDA & Distribution $\hat{P}_i(t)$ \\
		\midrule

		& \textbf{Diagnostics} \\
		DATO & OI, FSOI, Cov \\
		QMDA & Not provided \\
		\bottomrule
	\end{tabularx}
	\caption{Conceptual correspondences between DATO and QMDA.}
	\label{tab:corrispondenze}
\end{table}

\subsection{Offline phase comparison}
\label{sec:confronto_offline}

\paragraph{Kernel construction}
Both frameworks begin by constructing a kernel matrix from training data. DATO \cite{conti2025dato} uses a fixed-bandwidth isotropic Gaussian kernel evaluated on $m$ snapshots, at cost $O(nm^{2})$, while QMDA uses a variable-bandwidth kernel with bistochastic normalisation evaluated on $N$ samples, at cost $O(dN^{2})$:

\begin{align*}
  T_{\mathrm{kernel}}^{\mathrm{DATO}} &= O(n\cdot m^2), & & & T_{\mathrm{kernel}}^{\mathrm{QMDA}} &= O(d\cdot N^2).
\end{align*}

The two costs share the same algebraic structure --- quadratic in the number of training points, linear in the data-space dimension --- but are not directly comparable in practice, since $m \ne N$ in general. In the L63 benchmark ($m = 2\,800$, $N = 64\,000$, $n = d = 3$),

\[
  \frac{T_{\mathrm{kernel}}^{\mathrm{QMDA}}}{T_{\mathrm{kernel}}^{\mathrm{DATO}}} = \frac{N^2}{m^2} \approx 520,
\]

a factor that reflects an architectural choice rather than an inefficiency: the larger training set is required by the convergence regime $N \to \infty$ at fixed $L$ of Theorem~1 of \cite{giannakis2019qmda}.

\paragraph{Eigendecomposition} Here the two frameworks diverge in computational strategy. DATO solves a dense $m \times m$ generalised eigenvalue problem via Cholesky factorisation at cost $O(m^{3})$ or, for the $S$ dominant eigenpairs only, via Krylov methods at cost $O(m^{2}S)$. QMDA exploits the sparsity of its kernel matrix --- only $r$ nearest neighbours per row are retained --- and computes the $L$ leading eigenvectors via ARPACK at cost $O(LrN)$, with $r \ll N$ by construction ($r/N \approx 8\%$ in the L63 experiment). For matched sizes $m = N$ and $S = L$, QMDA's sparse Krylov strategy is therefore asymptotically cheaper than DATO's dense one whenever $r \ll m$.

\paragraph{Operator representations} The most consequential structural difference in the offline phase is whether the framework materialises the operator matrices explicitly. DATO does not: prediction is implicit, encoding the entire dynamics in the $S$ scalar eigenvalues $\{\lambda_{j}\}$, with no operator matrix ever assembled. QMDA, conversely, explicitly constructs the $L \times L$ Koopman matrix $U^{(q)}$ and all $S_{\text{QMDA}}$ spectral projectors $E_{i} \in \mathbb{R}^{L \times L}$, at total cost $O(N\ L^{2})$ in the base case of a single assimilation horizon $q$. This explicit materialisation is expensive offline but is precisely what makes the online cycle cost independent of $n$, $N$, and $q$ — a trade-off that becomes decisive at high state-space dimension.

\subsection{Online phase comparison}
\label{sec:confronto_online}

\paragraph{Prediction step}
\label{sec:conf_pred}
The prediction step is where the two frameworks diverge most sharply. In DATO, propagating the state from one assimilation cycle to the next reduces to $S$ scalar multiplications $\varepsilon^{(k+1)}_{j} = \lambda^{q}_{j} \varepsilon^{(k)}_{j}$, followed by a matrix--vector product of cost $O(mS)$ to reconstruct the prior density on the training points; the total cost $O(mS)$ is the minimum achievable for a basis of $S$ modes \citep[eq.~11--12]{conti2025dato}. In QMDA, the corresponding operation is a pair of $L \times L$ matrix--matrix products for the Schrödinger-picture evolution $\hat{\rho} = U^{(q)*}_{L} \hat{\rho}^{+} U^{(q)}_{L} / \text{tr}(\dots)$, at cost $O(L^{3})$. For equal mode counts $S = L$, the overload factor is

\[
  \frac{O(L^3)}{O(mL)} = \frac{L^2}{m},
\]

which for $L = 1\,000$ and $m = 2\,800$ evaluates to roughly $357$. The gap is a direct consequence of the implicit-versus-explicit representation: DATO encodes the dynamics in $S$ scalar eigenvalues, while QMDA propagates the density operator through the materialised matrix $U^{(q)}$.

\paragraph{Analysis step}
\label{sec:conf_analisi}

In DATO the analysis step applies a pointwise Gaussian likelihood on $m$ training points at cost $O(m(pn + p^{2}))$, reprojects the posterior onto the PF basis at cost $O(mS + S^{2})$, and reconstructs the analysis state as the posterior mean at cost $O(mn)$, for a total per-cycle cost $O(mn + mpn + mS + S^{2})$ that scales linearly in $n$ through the state reconstruction and the likelihood. In QMDA the update reduces to the projective operation $\hat{\rho}^{+} = E_{i}\hat{\rho}E_{i} / \text{tr}(\dots)$ at cost $O(L^{3})$, with no dependence on $n$. The comparison is summarised in \cref{tab:conf_analisi}.

\begin{table}[htp!]
	\centering
	\small
	\begin{tabularx}{\columnwidth}{@{}ll@{}}
		\toprule
		\textbf{Method} & \textbf{Cost per cycle} \\
		\midrule
		\textbf{DATO} \\ Likelihood, Bayes, PF proj. & $O(mn + mpn + mS + S^2)$ \\
		\midrule
		\textbf{QMDA} \\ 
		Proj. update $\mathbf{E}_i\hat{\rho}\mathbf{E}_i$ & $O(L^3)$ \\
		Meas.\ prob.\ ($S_{\mathrm{QMDA}}$ bins) & $O(S_{\mathrm{QMDA}}\cdot L)$ \\
		\bottomrule
	\end{tabularx}
	\caption{Comparison of the analysis step.}
	\label{tab:conf_analisi}
\end{table}

The relative advantage between the two frameworks depends on the application regime, governed by the magnitude of $n$ relative to the break-even threshold $L^{3}/m$, as summarised in \cref{tab:conf_regime}.

\begin{table}[htp!]
	\centering
	\small
	\begin{tabularx}{\columnwidth}{@{}llll@{}}
		\toprule
		\textbf{Regime} & \textbf{DATO} & \textbf{QMDA} & \textbf{Advantage} \\
		\midrule
		$n \ll L^3/m$ \\ (small state) & $O(mn)$  & $O(L^3)$ & DATO \\
		$n \gg L^3/m$ \\ (large state) & $O(mn)$  & $O(L^3)$ & QMDA \\
		$n = m = L$ & $O(n^2)$ & $O(n^3)$ & DATO \\
		$n \gg m \gg L$ \\ (Geophys. systems) & $O(mn)$ & $O(L^3)$ & QMDA \\
		\bottomrule
	\end{tabularx}
	\caption{Relative advantage in the analysis step across regimes. For DATO, the dominant term is $O(mpn)$ when $pn \gg S/m$, otherwise $O(mS+S^2)$.}
	\label{tab:conf_regime}
\end{table}

\paragraph{Dependence on the state dimension $n$}
\label{sec:conf_n}

The dependence of each computational phase on $n$, summarised in \cref{tab:conf_n}, is the most operationally significant structural difference between the two frameworks.

\begin{table}[htp!]
	\centering
	\small
	\begin{tabularx}{\columnwidth}{@{}lll@{}}
		\toprule
		\textbf{Phase} & \textbf{DATO} & \textbf{QMDA} \\
		\midrule
		Offline \\
		\quad kernel & $O(n\cdot m^2)$ & $O(d\cdot N^2)$ \\
		Online \\
		\quad prediction & $O(mS) = O(n^0)$  & $O(L^3) = O(n^0)$ \\
		\quad analysis & $O(mn) = O(n)$ & $O(L^3) = O(n^0)$ \\
		\quad state reconstr. & $O(mn) = O(n)$ & Not needed \\
		\bottomrule
	\end{tabularx}
	\caption{Dependence on the state-space dimension $n$ across computational phases.}
	\label{tab:conf_n}
\end{table}

The DATO online cycle scales as $O(n)$ per cycle through the likelihood evaluation and the state reconstruction, whereas the QMDA online cycle is entirely independent of the state dimension once the offline phase is complete: in the analysis step DATO costs $O(mn)$ against QMDA's $O(L^{3})$, regardless of $n$. For geophysical systems with $n \sim 10^{7}$--$10^{9}$ (global ocean or atmosphere models), this independence is decisive.

\paragraph{Break-even threshold and selection criterion}

Equating the total online costs of the two frameworks over $K$ cycles, for $S=L$, yields the break-even threshold:
\[
	n^* = \frac{L^3}{mp}
\]
if the number of  observations $p$ is fixed, with $p \ll m$, then the break-even threshold becomes:
\[
    n^* = \frac{L^3}{m}
\]

If $n < n^{*}$, the per-cycle cost of DATO is lower and the framework is more efficient overall; if $n > n^{*}$, the $n$-independence of QMDA's online cycle becomes decisive. The threshold depends only on $L$ and $m$ --- the multiplicative factor $K$ cancels in the ratio --- and for the L63 configuration ($L = 1\,000$, $m = 2\,800$) it evaluates to $n^{*} \approx 3.6 \times 10^{5}$, several orders of magnitude above the state dimension $n = 3$. The threshold becomes operationally relevant in high-dimensional geophysical applications, where $n$ comfortably exceeds $n^{*}$ for typical values of $L$ and $m$.

\begin{figure}[htb!]
	\centering
	\includegraphics[width=0.8\columnwidth]{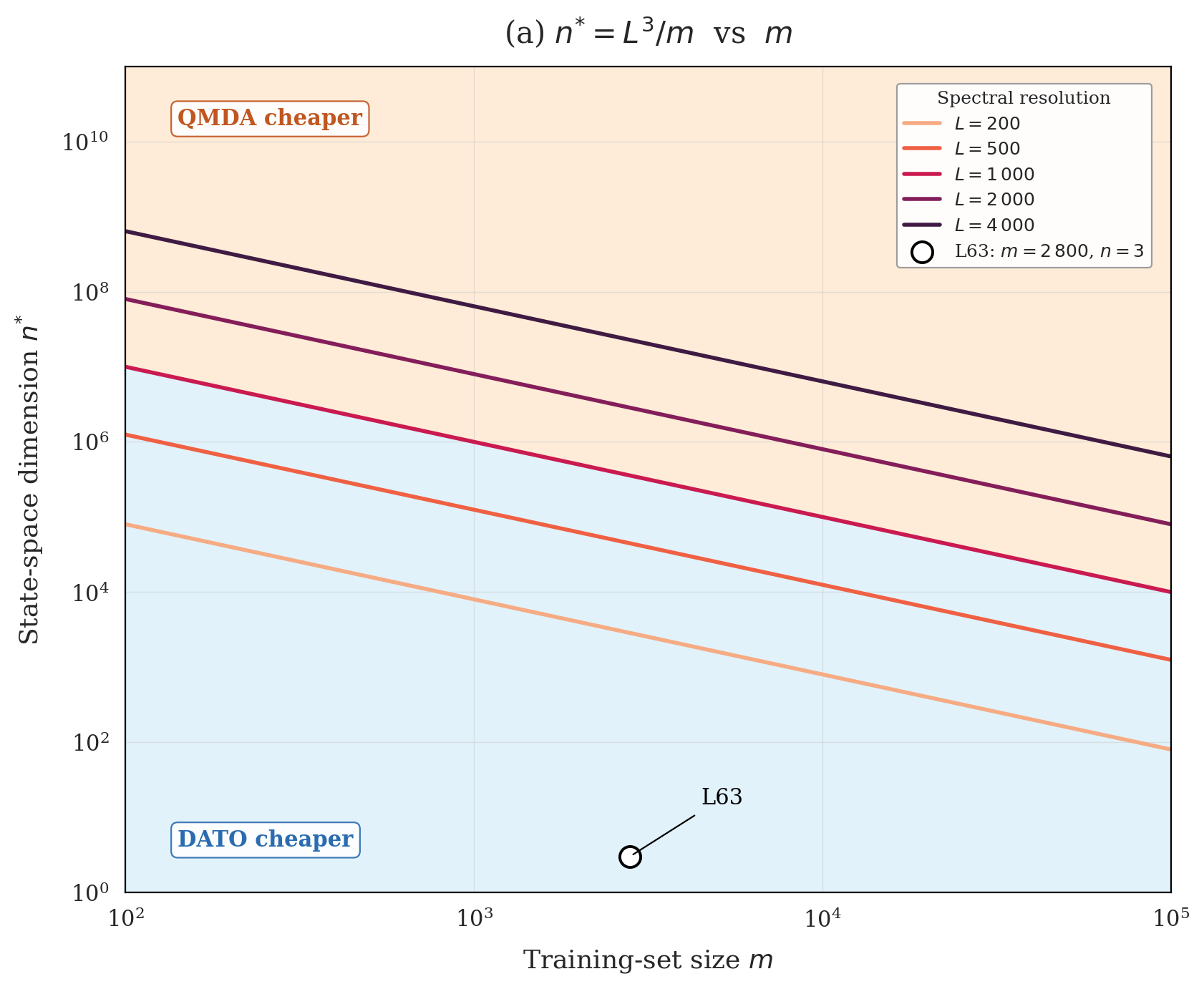}
	\includegraphics[width=0.8\columnwidth]{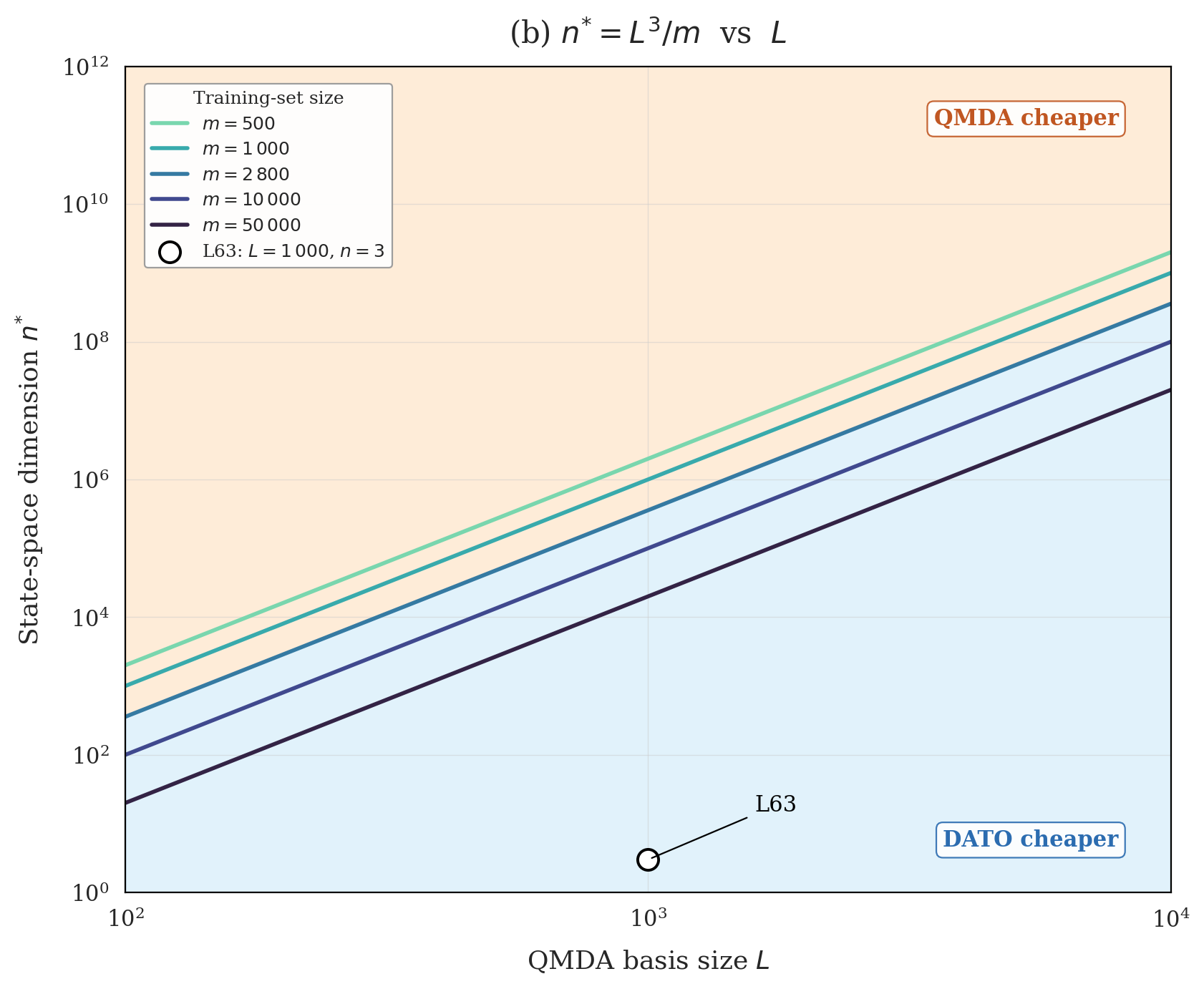}
	\includegraphics[width=0.8\columnwidth]{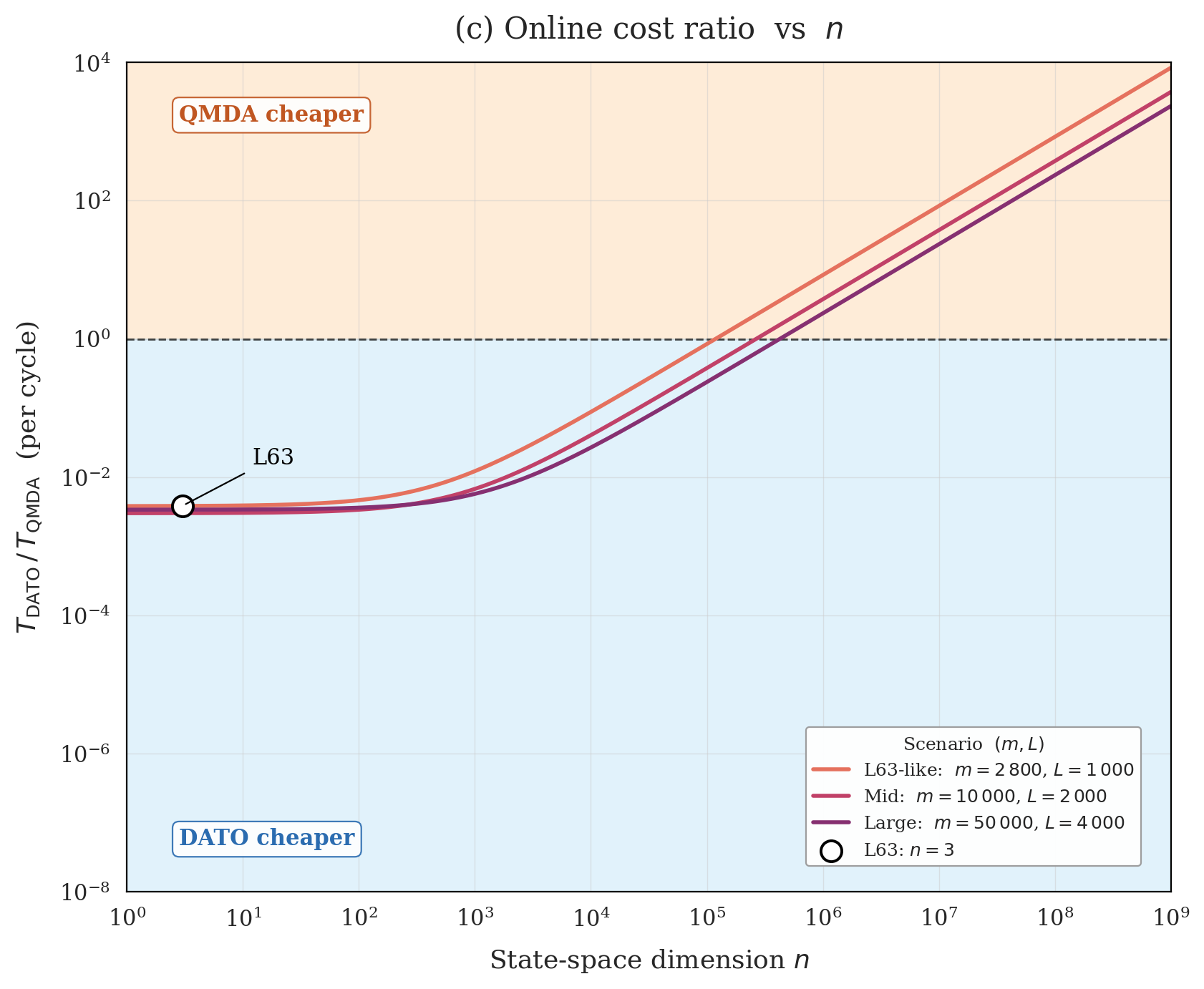}
	
	\caption{Three complementary views of the break-even threshold $n^{*} = L^{3}/m$: (a) as a function of the training-set size $m$, for several spectral resolutions $L$; (b) as a function of the QMDA basis size $L$, for several training-set sizes $m$; (c) the per-cycle online cost ratio $T_{\mathrm{DATO}}/T_{\mathrm{QMDA}}$ as a function of the state dimension $n$, for three representative configurations.}
	\label{fig:threshold_n}
\end{figure}

The three panels of \cref{fig:threshold_n} offer a consistent picture of the threshold. Panel (a) shows that $n^{*}$ decreases monotonically with $m$ at fixed $L$, while increasing $L$ shifts the family of curves upwards (cubically); panel (b) restates the same behaviour in dual form, with $n^{*}$ growing cubically in $L$ at fixed $m$ --- doubling $L$ multiplies $n^{*}$ by eight --- and a larger training set translating the curves downwards. Panel (c) translates the geometry into a per-cycle cost ratio: each curve is flat at small $n$, where DATO is dominated by the $n$-independent $S^{2}$ posterior projection, and transitions to a linear ramp once the $O(mn)$ term takes over and crosses the unit threshold at $n = n^{*}$, with the crossover shifting towards larger $n$ across the three scenarios (L63-like, mid, and large) as $n^{*} = L^{3}/m$ grows. In all three panels, the L63 configuration ($m = 2\,800$, $L = 1\,000$, $n = 3$, cost ratio $\approx 6 \times 10^{-3}$) lies firmly in the DATO-cheaper region, several orders of magnitude below the crossover.

\subsection{Qualitative dimensions}
\label{sec:riepilogo}

Beyond raw computational cost (\cref{tab:riepilogo}), the two frameworks differ along several qualitative dimensions that bear on the choice between them in practice. In terms of \textbf{output type}, DATO returns a point estimate $x^{a} \in \mathbb{R}^{n}$ together with a density on the training set, from which covariances and higher moments follow analytically; QMDA returns a discrete probability distribution over the partition bins, richer in distributional information but requiring post-processing to yield a state-space point estimate. As regards \textbf{diagnostics}, DATO admits closed-form observation influence (OI) and forecast sensitivity to observation impact (FSOI), which are traditionally available only through ensemble runs or adjoint models; QMDA provides no analogous diagnostics in its current formulation, although their derivation within the quantum-mechanical framework is identified as an open direction in \cite{giannakis2019qmda}. On the question of \textbf{convergence guarantees}, QMDA rests on Theorem~1 of \cite{giannakis2019qmda}, which establishes asymptotic consistency of the data-driven scheme in the limit $N \to \infty$ at fixed $L$, whereas DATO's convergence is established at the level of the kEDMD approximation, and a comparable theorem for the full assimilation cycle is still missing. Finally, both frameworks identify \textbf{quantum acceleration} of spectral decompositions as a natural future direction, with the operator-theoretic structure of QMDA making it a particularly natural candidate for implementation on quantum hardware.

\begin{table}[htp!]
	\centering
	\small
	\begin{tabularx}{\columnwidth}{@{}lll@{}}
		\toprule
		\textbf{Offline} \\
		& \textbf{Total (base case)} \\
		DATO & $O(nm^2 + m^3)$ \\
		QMDA & $O(dN^2 + NL^2)$ \\
		\midrule
		\textbf{Online (per cycle)} \\
		& \textbf{Prediction} \\
		DATO	& $O(S) + O(mS)$ \\
		QMDA	& $O(L^3)$ \\
		& \textbf{Measure probability} \\
		DATO	& (continuous likelihood) \\
		QMDA	& $O(S_{\mathrm{QMDA}}\cdot L)$ \\
		& \textbf{Analysis} \\
		DATO	& $O(mn + mpn + mS + S^2)$ \\
		QMDA	& $O(L^3)$ \\
		\midrule
		\textbf{Dependence on} \\
		& \textbf{$\boldsymbol{n}$ (online)} \\
		DATO	& $O(n)$ per cycle \\
		QMDA	& $O(n^0) = O(1)$ \\
		& \textbf{$\boldsymbol{m/N}$ (online)} \\
		DATO	& $O(m)$ per cycle \\
		QMDA	& $O(m^0) = O(1)$ \\
		\midrule
		\textbf{Memory (total)} \\
		DATO	& $O(m^2 + mn)$ \\
		QMDA	& $O(NL + S_{\mathrm{QMDA}}L^2)$ \\
		\midrule
		\textbf{Eigendecomposition} \\
		DATO	& $O(m^3)$ or $O(m^2 S)$ \\
		QMDA	& $O(L\cdot r\cdot N)$ (sparse) \\
		\bottomrule
	\end{tabularx}
	\caption{Comparative summary of the computational complexity of DATO and QMDA.}
	\label{tab:riepilogo}
\end{table}

\subsection{Preference scenarios}
\label{sec:scenari}

The qualitative comparison of \cref{sec:riepilogo} suggests a set of practical guidelines, though the decision typically involves weighing several considerations simultaneously.

DATO is the more natural choice when the state dimension satisfies $n \lesssim L^{3}/m$, since in this regime its per-cycle burden $O(mn)$ stays comfortably below QMDA's fixed $O(L^{3})$. It is particularly attractive when a point estimate $x^{a}\in\mathbb{R}^{n}$ is required in state space, when the Gaussian observation likelihood \eqref{eq:cost_lik} is correctly specified, and when understanding how individual observations shape the analysis is itself a scientific priority --- the closed-form OI and FSOI diagnostics, available as a by-product of the operator-theoretic structure, replacing ensemble or adjoint computations that would otherwise be needed. Long assimilation horizons combined with an effective spectral truncation $S \ll m$ further strengthen the case, reducing the per-cycle cost to $O(mS)$.

QMDA becomes the more compelling option as $n$ grows beyond $L^{3}/m$, where the $n$-independence of its online step turns into a decisive structural advantage. It is also the natural choice whenever the full posterior, rather than a point estimate, is the primary output of interest: the density operator $\hat{\rho}$ encodes the complete posterior over the partition bins, a representation particularly valuable in non-Gaussian or multimodal regimes where collapsing to a posterior mean would discard physically relevant information. The asymptotic-consistency guarantees of \cite[Theorem~1]{giannakis2019qmda} lend further theoretical robustness, conditional on the observed variable admitting a coarse-graining into $S$ bins without unacceptable information loss. The operator-algebraic structure of QMDA finally positions it as a natural candidate for quantum-hardware implementation, an avenue along which \cite{freeman2023} has begun to elaborate the theoretical underpinnings.

%% file: sec_6_lorenz_63.tex
\section{Practical estimates on the Lorenz--63 benchmark}
\label{sec:l63}

The L63 system introduced in \cref{sec:introduzione} provides the natural setting in which to translate the asymptotic profiles of the previous sections into concrete operational figures. Its small state dimension makes the benchmark a regime in which neither framework is computationally stressed by the physics, so that the estimates below serve as a consistency check between the asymptotic analysis and the orders of magnitude observed in published experiments, rather than as a stress test of either method.

\paragraph{DATO on Lorenz--63}

The configuration adopted by \citet{conti2025dato} integrates the system on $t \in [0, 100]$ with a fourth-order Runge--Kutta scheme at time step $\Delta t = 0.025$, discards the first $30\%$ of the trajectory as transient, and retains the remaining $70\%$ as the training set, yielding $m \approx 2\,800$ snapshots \citep[Sec. 4a]{conti2025dato}. The kernel-EDMD step employs a Gaussian RBF kernel with bandwidth $\sigma_{\mathrm{kernel}} = 2$ and Tikhonov regularisation $\varepsilon = 10^{-5}$, and the leading $S = 2\,000$ eigenpairs are retained for the spectral representation. The assimilation experiment is run on $t \in [0, 60]$, with observations of the $y$ and $z$ components only ($p = 2$) collected every six model steps and Gaussian observation noise of standard deviation $\sigma_{\mathrm{obs}} = 0.5$. The resulting operational estimates, reported in \cref{tab:dato_l63}, confirm the regime predicted by the asymptotic analysis.

\begin{table}[htp!]
  \centering
  \small
  \begin{tabularx}{\columnwidth}{@{}Xr@{}}
    \toprule
    \textbf{Phase} & \textbf{Estimated operations} \\
    \midrule
    \multicolumn{2}{@{}l}{\textbf{Offline}} \\
    \quad Gram matrix              & $O(nm^{2}) \approx 2.4\times10^{7}$ \\
    \quad Eigendecomp.\ (Krylov)   & $O(m^{2}S) \approx 1.6\times10^{10}$ \\
    \\
    \multicolumn{2}{@{}l}{\textbf{Online (per cycle)}} \\
    \quad Prediction               & $O(mS) \approx 5.6\times10^{6}$ \\
    \quad Likelihood               & $O(mpn) \approx 1.7\times10^{4}$ \\
    \quad Projection               & $O(mS+S^{2}) \approx 9.6\times10^{6}$ \\
    \quad State reconstruction     & $O(mn) \approx 8.4\times10^{3}$ \\
    \midrule
    \textbf{Offline bottleneck}    & Eigendecomp.\ $O(m^{2}S)$ \\
    \textbf{Online bottleneck/cyc.}& Projection $O(mS+S^{2})$ \\
    \bottomrule
  \end{tabularx}
  \caption{Operational estimates for the DATO offline and online phases on L63. Parameters: $n=3$, $m=2\,800$, $S=2\,000$, $p=2$.}
  \label{tab:dato_l63}
\end{table}

The offline phase is dominated, in this configuration, by the eigendecomposition step, whose cost $O(m^{2}\,S) \approx 1.6 \times 10^{10}$ exceeds that of the Gram-matrix construction by three orders of magnitude. Within the online cycle, the bottleneck is not the likelihood evaluation, which contributes only $O(m\,p\,n) \approx 1.7 \times 10^{4}$ operations per cycle owing to the small observation and state dimensions, but the posterior projection onto the PF basis, whose $O(m\,S + S^{2}) \approx 9.6 \times 10^{6}$ operations exceed every other contribution; the projection cost is, in turn, almost equally split between its two terms ($m\,S = 5.6 \times 10^{6}$ and $S^{2} = 4.0 \times 10^{6}$), in agreement with the $S/m \approx 0.71$ ratio adopted by \citet{conti2025dato}, which is precisely the regime in which the $S^{2}$ term cannot be neglected against $m\,S$ and in which the projection takes over from the likelihood as the dominant per-cycle operation.

\paragraph{QMDA on Lorenz--63}

The configuration adopted by \citet{giannakis2019qmda} is qualitatively different. The training set is much larger, with $N = 64\,000$ samples drawn from the L63 trajectory, and the kernel matrix is replaced by its sparse approximation retaining $r = 5\,000$ nearest neighbours per row --- approximately $8\%$ of $N$. The eigendecomposition retains $L = 1\,000$ leading eigenvectors of the sparse kernel, the observation partition has $S_{\mathrm{QMDA}} = 32$ bins, and the maximum lag for the precomputed Koopman matrices is $q = 100$ time steps, the value used to produce the multi-horizon outputs of Figures~5 and~6 of \citet[Appendix~B and Sec.~VI]{giannakis2019qmda}. The corresponding operational estimates are reported in \cref{tab:qmda_l63}.

\begin{table}[htp!]
  \centering
  \small
  \begin{tabularx}{\columnwidth}{@{}Xr@{}}
    \toprule
    \textbf{Phase} & \textbf{Estimated operations} \\
    \midrule
    \multicolumn{2}{@{}l}{\textbf{Offline}} \\
    \quad Kernel matrix                 & $O(dN^{2}) \approx 1.2\times10^{10}$ \\
    \quad Eigenvectors (ARPACK)         & $O(LrN) \approx 3.2\times10^{11}$ \\
    \quad Koopman $U^{(q)}$ (single)    & $O(NL^{2}) \approx 6.4\times10^{10}$ \\
    \quad Koopman $U^{(0)},\dots,U^{(q)}$ (multi-horizon) & $q\,O(NL^{2}) \approx 6.4\times10^{12}$ \\
    \\
	\multicolumn{2}{@{}l}{\textbf{Online (per cycle)}} \\
    \quad Evolution                     & $O(L^{3}) = 10^{9}$ \\
    \quad Meas.\ probabilities          & $O(S_{\mathrm{QMDA}}L) \approx 3.2\times10^{4}$ \\
    \quad Projective update             & $O(L^{3}) = 10^{9}$ \\
    \midrule
    \textbf{Offline bottleneck}         & Eigenvectors $O(LrN)$ \\
    \textbf{Online bottleneck}          & Evol.\ + update $O(L^{3})$ \\
    \bottomrule
  \end{tabularx}
  \caption{Operational estimates for the QMDA offline and online phases on L63. Parameters: $d=3$, $N=64\,000$, $L=1\,000$, $r=5\,000$, $S_{\mathrm{QMDA}}=32$, $q=100$.}
  \label{tab:qmda_l63}
\end{table}

As anticipated by the asymptotic analysis, the offline phase is dominated by the kernel-eigenvector computation $O(L\,r\,N) \approx 3.2 \times 10^{11}$, ahead of the kernel-matrix assembly $O(d\,N^{2}) \approx 1.2 \times 10^{10}$ and of the construction of each operator matrix $O(N\,L^{2}) \approx 6.4 \times 10^{10}$. The choice of precomputing the full set $U^{(0)}, \ldots, U^{(100)}$ of Koopman matrices, however, multiplies the latter contribution by a factor of $101$, raising the offline cost to roughly $6.4 \times 10^{12}$ operations and making the multi-horizon precomputation, rather than the kernel itself, the most expensive step in this specific setting. The online cycle exhibits the dimension-independent $O(L^{3}) = 10^{9}$ cost characteristic of QMDA, against which the measurement-probability contribution $O(S_{\mathrm{QMDA}}\,L) \approx 3.2 \times 10^{4}$ is negligible: per-cycle work is concentrated in the two pairs of $L \times L$ matrix--matrix products of the unitary evolution and of the projective update.
The Lorenz--63 estimates confirm, on a common benchmark, the qualitative picture drawn by the asymptotic analysis: DATO operates in a regime whose per-cycle bottleneck is the posterior projection $O(m\,S + S^{2})$, whereas QMDA operates at a fixed online cost $O(L^{3})$ pre-paid through a substantially larger offline phase. The two configurations differ by more than two orders of magnitude in their per-cycle cost despite sharing the same physical state dimension $n = 3$, a quantitative reminder that the choice between the frameworks reflects qualitatively different scaling laws rather than constant factors.

%% file: sec_7_conclusions.tex
\section{Conclusions}
\label{sec:conclusioni}

This work has placed DATO and QMDA on a uniform analytical footing and derived, for each framework, a complete account of its offline and online complexity. Despite the shared Koopman-operator foundation, the two algorithms occupy structurally different regions of the cost landscape, and the analysis has identified in quantitative terms the regimes in which one is preferable to the other.
The principal outcome of the comparison is the break-even threshold derived in \Cref{sec:confronto} under the natural identification $S \approx L$,
\[
  n^{*} \;=\; \frac{L^{3}}{m},
\]
which separates the regime $n < n^{*}$, where the linear $n$-dependence of the DATO online cycle is offset by its lower offline footprint, from the regime $n > n^{*}$, in which the $n$-independence of the QMDA cycle becomes the decisive structural advantage. The two profiles are nevertheless not exhausted by their per-cycle costs: DATO collapses to either $O(mpn)$ or $O(mS + S^{2})$ depending on the regime, while QMDA collapses every per-cycle operation to $O(L^{3})$ at the price of an offline phase dominated by the kernel construction $O(d\,N^{2})$ and a memory footprint $O(NL + S_{\mathrm{QMDA}}\,L^{2})$ that scales bilinearly, rather than quadratically as the $O(m^{2})$ Gram matrices of DATO. The natural outputs --- a point estimate $x^{a}$ with closed-form OI and FSOI diagnostics in DATO, a discrete distribution $\hat{P}_{i}(t)$ over the observable partition in QMDA --- constitute a structural difference that, in practice, often weighs on the choice as heavily as the asymptotic cost itself.

\paragraph{Limitations of the analysis.}
Two caveats deserve mention. First, the complexity analysis is carried out in worst-case big-$O$ form and abstracts from hidden constants, BLAS/LAPACK details, cache behaviour, parallelism, and the spectral-gap dependence of iterative solvers; the operational estimates of \cref{sec:l63} and the threshold $n^{*}$ should accordingly be read as structural rather than operationally tight criteria. Second, the QMDA analysis is restricted to the base implementation in which a single Koopman matrix $U^{(q)}$ is precomputed: the alternative strategy based on the matrix power $(U^{(1)})^{q}$ shifts cost from offline to online and is left for future quantitative comparison.

\section{Future works}
\label{sec:future_works}

The analysis suggests several natural follow-up directions. The most immediate is the empirical validation of the asymptotic profiles, through a systematic timing study on L63 and on a synthetic family of systems of increasing state dimension $n$, to estimate the hidden constants of the big-$O$ bounds and to localise $n^{*}$ in practice. A complementary step is the assessment of \textbf{scalability on real hardware}, by porting both frameworks to HPC environments and quantifying their behaviour under multi-core/GPU parallelism, cache hierarchy, and distributed-memory communication --- ingredients that the asymptotic analysis deliberately abstracts away. A natural extension is the migration of the comparison to genuinely high-dimensional systems (Lorenz~96, quasi-geostrophic, or shallow-water models), in which the structural advantage of QMDA in the large-$n$ regime should become empirically observable.

On the methodological side, the design of hybrid frameworks that combine the implicit eigenvalue-based representation of DATO with the full distributional output of QMDA appears particularly promising, as does the extension of both schemes to non-stationary dynamics through time-dependent kernels or online updates of the spectral basis. Finally, the structural similarity between QMDA and the formalism of quantum mechanics, jointly emphasised by \citet{conti2025dato} and \citet{giannakis2019qmda}, makes quantum acceleration of the offline spectral decompositions a natural longer-term research line, whose impact on $n^{*}$ would be worth quantifying within a unified benchmark for DATO, QMDA, and their hybrid extensions. As stressed by \citet{freeman2023}, however, porting QMDA to a genuine quantum environment introduces challenges of its own --- the iterative forecast--analysis cycle requires repeated interaction between quantum hardware and the classical assimilated system, possibly mediated by quantum sensors --- and the design of efficient interfaces for this hybrid quantum--classical loop constitutes a fruitful research direction in itself.

%% file: sec_8_CRediT.tex
\section*{CRediT authorship contribution statement}
\textbf{Emanuele Donno}: Conceptualization, Formal Analysis, Investigation, Methodology, Writing -- original draft, Writing -- review \& editing.
\textbf{Giovanni Conti}: Conceptualization, Validation, Writing -- review \& editing.
\textbf{Paolo Oddo}: Validation.
\textbf{Silvio Gualdi}: Validation.
\textbf{Luca Mainetti}: Validation.
\textbf{Giovanni Aloisio}: Supervision, Conceptualization, Validation, Writing -- review \& editing.